\documentclass[useAMS,usenatbib]{mn2e}
\usepackage{epsfig}
\usepackage{amsmath}
\usepackage{rotating}

\newcommand{\be}{\begin{equation}}
\newcommand{\beq}{\begin{equation}}
\newcommand{\ba}{\begin{eqnarray}}
\newcommand{\ee}{\end{equation}}
\newcommand{\eeq}{\end{equation}}
\newcommand{\ea}{\end{eqnarray}}
\newcommand{\msun}{M_{\odot}\hspace{1mm}}

\def\lsim{~\rlap{$<$}{\lower 1.0ex\hbox{$\sim$}}}

\def\gsim{~\rlap{$>$}{\lower 1.0ex\hbox{$\sim$}}}

\title[SZ vs. X--ray Scaling Relation]{Probing Cosmology and Galaxy Cluster Structure with the Sunyaev--Zel'dovich Decrement vs. X--ray Temperature Scaling Relation}
\author[Shang et al.]{Cien Shang$^{1}$, Zolt\'an Haiman$^{2}$, Licia Verde$^{3,4}$\\
$^1$Department of Physics, Columbia University, 550 West 120th Street, New York, NY 10027, USA; cien@phys.columbia.edu\\
$^2$Department of Astronomy, Columbia University, 550 West 120th Street, New York, NY 10027, USA; zoltan@astro.columbia.edu\\
$^3$Institute of Space Sciences (CSIC-IEEC), UAB, Barcelona 08193, Spain; verde@ieec.uab.es\\
$^4$Department of Astrophysical Sciences, Princeton University, Ivy Lane, Princeton, USA}
\begin{document}

\date{\today}
\pagerange{\pageref{firstpage}--\pageref{lastpage}} \pubyear{2008}

\maketitle

\label{firstpage}

\begin{abstract}
Scaling relations among galaxy cluster observables, which will become
available in large future samples of galaxy clusters, could be used to
constrain not only cluster structure, but also cosmology.  We study
the utility of this approach, employing a physically motivated
parametric model to describe cluster structure, and applying it to the
expected relation between the Sunyaev-Zel'dovich decrement ($S_{\nu}$)
and the emission--weighted X--ray temperature ($T_{\mathrm{ew}}$). The
slope and normalization of the entropy profile, the concentration of
the dark matter potential, the pressure at the virial radius, and the
level of non-thermal pressure support, as well as the mass and
redshift--dependence of these quantities are described by free
parameters. With a suitable choice of fiducial parameter values, the
cluster model satisfies several existing observational constraints.
We employ a Fisher matrix approach to estimate the joint errors on cosmological
and cluster structure parameters from a measurement of $S_{\nu}$ {\it
vs.}  $T_{\mathrm{ew}}$ in a future survey.  We find that different
cosmological parameters affect the scaling relation differently:
predominantly through the baryon fraction ($\Omega_m$ and $\Omega_b$),
the virial over--density ($w_0$ and $w_a$ for low--$z$ clusters) and
the angular diameter distance ($w_0$, $w_a$ for high--$z$ clusters;
$\Omega_{DE}$ and $h$).  We find that the cosmology constraints from
the scaling relation are comparable to those expected from the number
counts ($dN/dz$) of the same clusters.  The scaling relation approach
is relatively insensitive to selection effects and it offers a
valuable consistency check; combining the information from the scaling
relation and $dN/dz$ is also useful to break parameter degeneracies and 
help disentangle cluster physics from cosmology.
Our work suggests that scaling relations should be a useful component
in extracting cosmological information from large future cluster
surveys.
\end{abstract}

%%%%%%%%%%%%%%%%%%%%%%%%%%%%%%%%%%%%%%%%%%%%%%%%%%%%%%%%%%%%%%%%%%%%%%%%%%%%%%%%
\section{Introduction}
\label{sec:introduction}
%%%%%%%%%%%%%%%%%%%%%%%%%%%%%%%%%%%%%%%%%%%%%%%%%%%%%%%%%%%%%%%%%%%%%%%%%%%%%%%%

This work is motivated by large upcoming cluster surveys that utilize
the Sunyaev-Zel'dovich (SZ) effect (Sunyaev \& Zeldovich 1972) such as
ACT\footnote{http://www.physics.princeton.edu/act/},
APEX\footnote{http://bolo.berkeley.edu/apexsz/},
Planck\footnote{http://www.rssd.esa.int/index.php?project=Planck}, and
SPT\footnote{http://pole.uchicago.edu/}. As is well known, the SZ
signal is nearly redshift independent, so these surveys are expected
to be especially efficient in detecting high--redshift clusters.  The
expected catalogs will be sensitive probes of dark energy, and also
useful in breaking degeneracies in local cluster surveys (for example,
the degeneracy between $\sigma_8$ and $\Omega_m$). The planned and
on--going surveys will cover thousands of square degrees of sky, and
detect on the order of $\sim$ 10,000 of clusters with masses over a
few $10^{14} {\rm M_\odot}$. For example, the SPT survey will cover
4,000 $\mathrm{deg}^2$ of sky in 4 frequency channels (90, 150, 220,
270 GHz), and Planck aims to cover the whole sky in 9 frequency
channels. These cluster samples will contain a significant amount of
cosmological information.

Importantly, cosmological information can be extracted from large
galaxy cluster catalogs in several complementary ways. For example,
the cluster abundance is exponentially sensitive to the amplitude of
matter density fluctuations, and the X-ray temperature function
obtained from local cluster samples has been used to constrain
$\Omega_m$ and $\sigma_8$ (e.g., Henry 2000, Ikebe et al. 2002). The
redshift evolution of the abundance will be useful in constraining
dark energy parameters, with statistical errors competitive with those
in most other methods (e.g., Haiman et al. 2001; Albrecht et
al. 2006). Small existing samples of tens of X--ray clusters out to
$z\sim 0.5$ already provide interesting constraints on the dark
energy density $\Omega_{DE}$ and equation of state parameter $w$
(Henry 2004; Mantz et al. 2008; Vikhlinin et al. 2008). The cluster
power spectrum also contains information on cosmology (Hu \& Haiman
2003), both through the growth of fluctuations (Refregier et al. 2002)
and through baryon acoustic features (Hu \& Haiman 2003; Blake \&
Glazebrook 2003, Seo \& Eisenstein 2003, Linder 2003). Combining the
number counts and the power spectrum provides a cross-check and can
allow a ``self--calibration'' to contain systematic errors in the
mass--observable relation (Majumdar \& Mohr 2004; Wang et
al. 2004). In addition to the above, clusters could also be used as
``standard rulers''. The measured gas fraction $f_{\mathrm{gas}}$,
which is derived from the observed X-ray temperature and density
profiles, depends on the angular diameter distance as
$f_{\mathrm{gas}} \propto D_A^{1.5}$ (e.g., Allen et al. 2008).  To the
extent that the gas fraction is predictable ab--initio from numerical
simulations, this provides a measurement of $D_A(z)$. A complementary
measurement of $D_A(z)$ can be provided by combining SZ and X-ray
signals, under the assumption that clusters are at least statistically
spherical (e.g., Bonamente et al. 2006).

The gravitational potential of clusters is dominated by dark matter,
whose behavior is determined by gravity alone, and is therefore
robustly predictable (Navarro et al. 1997; hereafter NFW).  If
astrophysical processes in the gas were unimportant, the intracluster
gas would evolve adiabatically, tracing the self--similar dark matter
profile, and its global properties would obey simple scaling relations
(e.g., Kaiser 1986).  In fact, observed clusters indeed exhibit scaling
relations that are tight, but which deviate significantly from the
self--similar expectation.  For example, the relation between X--ray
flux ($L_X$) and temperature ($T_X$) is observed to be close to $L_X
\propto T_X^3$ , significantly steeper than the $L_X \propto T_X^2$
power law expected in self-similar, adiabatic models.  These
observations could be explained by preferentially increasing the
specific entropy of the cluster gas in low--mass clusters. Many
variants of such models have been developed, based either on heat
input from stars or nuclear black holes, or preferential elimination
of the low--entropy gas by star--formation (see, e.g., a review by
Voit 2005 and references therein).  Our present study is motivated by
the fact that in any such model, the predicted scaling relations will
generically depend on the background cosmology.  Using simple toy
models, Verde et al (2002; hereafter VHS02) showed that the
cosmological parameters indeed affect cluster scaling relations,
i.e. relations among temperature, cluster size and SZ decrement.  In
small cluster samples (e.g., Morandi et al. 2007), the subtle
cosmology--dependencies will be masked by the larger uncertainties in
the physical modeling of cluster structure.  However, given a
sufficiently accurate measurement of the scaling relations, using
thousands of clusters, it should become possible to place useful
constraints simultaneously on cosmological parameters and the
parameters of any given specific cluster physics model.

VHS02 argued that combining SZ and X--ray data will be particularly
useful, because the SZ and X--ray signatures depend on cosmological
parameters differently, and singled out the relation between the
Sunyaev-Zel'dovich decrement ($S_{\nu}$) and the X--ray temperature
($T_X$) as a promising probe of both cluster structure and cosmology.
Afshordi (2008) showed that the measured relation between SZ decrement
and angular half--light radius, which does not require X--ray data,
may already help reduce the errors in cluster mass estimates. Younger
et al. (2006) showed that combining number counts from SZ and X-ray
surveys delivers constraints that are tighter than adding two
independent measurements in quadrature; this synergy again arises
because the SZ decrement and X--ray flux depend differently on the
background cosmology.  Finally, Aghanim et al. (2008) recently used
hydrodynamical simulations, and studied how different values of the
dark energy equation of state $w$ affect SZ vs. X--ray scaling
relations. They found relatively little direct sensitivity to $w$
(which is consistent with our own findings; see discussion in
\S~\ref{subsec:results_c} below).

Despite the above few works, the utility of the scaling relations in
probing cosmology remains relatively unexplored. We believe it
deserves more investigation, for the following two reasons.  First,
data on the scaling relations will be automatically available (at
least for a subset of clusters) once the planned SZ surveys are
performed. Large catalogs of cluster temperatures (hundreds of
clusters) already exist, and new, much deeper X-ray surveys are being
proposed and planned, such as eROSITA and IXO\footnote{See http://www.mpe.mpg.de/erosita/MDD-6.pdf and http://ixo.gsfc.nasa.gov, respectively.}.
Compared to the number counts, the scaling relation technique should
be relatively less sensitive both to selection effects and to the
relation between the observables and cluster mass.  Second, as we will
discuss below in detail, scaling relations derive cosmological
information from a different combination of geometrical distances and
non--linear growth than the other cluster observables. For this
reason, they could not only be combined with other techniques to
tighten constraints, but can also serve as useful consistency checks.

In this paper, we follow VHS02, and we focus on the relation between
the total SZ flux decrement, encoded in integrated Compton $y$
parameter, and the X-ray emission weighted temperature. There are
other physical quantities, such as the X-ray luminosity or the central
SZ decrement $y_0$.  These quantities are especially sensitive to the
properties of the cluster core, where cooling, star--formation, and
feedback processes are most effective, and which is therefore the most
difficult region of the cluster to model.  The scatter in these
quantities is known to be large, which will limit their
utility for constraining cosmology.  In contrast, the integrated
Compton $y$ parameter and the mean emission--weighted temperature show
strong robustness to the above uncertainties (Reid \& Spergel 2006 and
Kravtsov, Vikhlinin \& Nagai 2006 showed that similarly robust
observables can be constructed from X--ray data, as well).  An
additional virtue of these two quantities is that they are relatively
easy to measure, i.e. they do not require a detailed measurement of
radial profiles.\footnote{We note, however, that the cluster core
needs to be excised in cooling core clusters, in order not to affect
the emission--weighted ICM temperature measurement. While it is
possible to extract both the core temperature and the ICM temperature
from a single spectrum, this will inevitably introduce uncertainties,
which will be discussed in \S~\ref{sec:discussion} below.}
The main improvements of the
present study over the analysis of VHS02 are the following: (i) we
include a full set of 8 cosmological parameters, representing the
matter density $\Omega_{m}$, the dark energy density $\Omega_{DE}$,
the baryon density $\Omega_{b}$, the Hubble constant $h$ ($h\equiv
H/100 \mbox{km s}^{-1} \mathrm{Mpc^{-1}}$), two dark energy equation
of state parameters $w_0$ and $w_a$ (see detailed definition in next
section), the normalization of the matter density power spectrum
$\sigma_8$ and the ``tilt'' of the primordial power spectrum $n_s$.
Note that we do not assume spatial flatness, so the 8 parameters are
independent. VHS02 only included $\Omega_{m}$, $\sigma_8$ and $h$ as
free parameters. (ii) VHS02 adopted a simple spherical toy model for
cluster structure, based on the virial theorem, to predict relations
between different observable quantities. This approach has the virtue
of simplicity, and makes it easier to interpret the results; however,
such a simple cluster model is already in contradiction with existing
data.  Here we use a more elaborate, and more realistic
phenomenological cluster model, with many free parameters.  We
explicitly require the model to satisfy existing observational
constraints and we explore the impact of various cluster structure
uncertainties on the final conclusions. (iii) We employ a Fisher
matrix technique, instead of a Kolmogorov-Smirnov test as in
VHS02. The Fisher matrix technique is a fast way of estimating joint
parameter uncertainties in a multi--dimensional parameter space, and
allows us to understand parameter degeneracies. (iv) Finally, we also
study constraints from the number counts (including the effects of
cluster structure uncertainty, mass-observable scatter and
incompleteness), and we forecast the combined constraints from the
scaling relations and the number counts.

The rest of this paper is organized as follows. In \S~\ref{sec:model},
we describe the Fisher matrix technique and the physically motivated,
phenomenological cluster model we adopt. The cluster model is compared
against observations and simulations. We also explain our choice of
fiducial values for cosmological parameters, cluster parameters and
survey parameters. In \S~\ref{sec:results}, we present our main
results, i.e. the constraints on cosmological and cluster structure
parameters.  Proceeding pedagogically, we first include only the 8
cosmological parameters, then add increasing uncertainties from the
cluster structure parameters to our analysis. We also explain in
detail where the cosmological constraints from the scaling relations
come from. In \S~\ref{sec:discussion}, we compare the scaling relation
technique with constraints from the number counts, and discuss various
caveats and possible improvements to our results. We summarize our
results and offer our conclusions in \S~\ref{sec:conclusion}.

%\vspace{3\baselineskip}

%%%%%%%%%%%%%%%%%%%%%%%%%%%%%%%%%%%%%%%%%%%%%%%%%%%%%%%%%%%%%%%%%%%%%%%%%%%%%%%%
\section{Cluster model and Fisher matrix technique}
\label{sec:model}
%%%%%%%%%%%%%%%%%%%%%%%%%%%%%%%%%%%%%%%%%%%%%%%%%%%%%%%%%%%%%%%%%%%%%%%%%%%%%%%%

%%%%%%%%%%%%%%%%%%%%%%%%%%%%%%%%%%%%%%%%%%%%%%%%%%%%%%%%%%%%%%%%%%%%%%%%%%%%%%%%
\subsection{Fisher matrix technique}
\label{subsec:fisher}
%%%%%%%%%%%%%%%%%%%%%%%%%%%%%%%%%%%%%%%%%%%%%%%%%%%%%%%%%%%%%%%%%%%%%%%%%%%%%%%%

We employ the Fisher matrix technique to forecast cosmological
constraints from future surveys. The Fisher matrix is a quick way to
estimate joint parameter uncertainties in a multi-parameter fit
(Fisher 1935; Tegmark et al. 1997). It is defined as,
\begin{eqnarray}
F_{ij}=-\left<\frac{\partial^{2}\ln{\cal{L}}}{\partial p_i \partial p_j}\right>,
\label{eqn:fisher}
\end{eqnarray}
where $\cal{L}$ is the likelihood for a certain observable, and $p_i$
is the parameter set (including both cosmological and cluster
structure parameters in our case). The best attainable covariance
matrix $C$ is simply the inverse of the Fisher matrix $F$,
\begin{eqnarray}
C_{ij}=\left(F^{-1}\right)_{ij},
\end{eqnarray}
and the constraint on any individual parameter $p_i$, marginalized
over all other parameters, is $\sqrt{(F^{-1})_{ii}}$. Another
advantage of the Fisher matrix technique is that it is easy to obtain
joint constraints from several data sets or methods: the total Fisher
matrix is just the sum of individual Fisher matrices as long as they
are uncorrelated. In this paper, we assume that the different Fisher matrices
are indeed independent; we justify this assumption in \S~\ref{sec:discussion}. 

The Fisher matrix approach makes the underlying assumption that
the likelihood surface for the parameters is a multi-variate Gaussian.
This is indeed the case if experimental errors are
Gaussian--distributed and the model depends linearly on the
parameters, but in general, this assumption does not hold, and is
instead justified by invoking the central limit theorem in the
presence of large number of independent data.  The classical example
is the CMB likelihood which is very close to Gaussian for the
so-called ``normal" or ``physical" parameters (Kosowsky et al. 2002),
but not necessarily for the standard cosmological parameters. However,
for most cosmological models and future CMB data sets, especially if
combined with external datasets or a weak prior on $H_0$, the CMB
likelihood is very close to Gaussian even for the standard
cosmological parameters (Komatsu et al. 2009).  For degeneracies in
parameters space that are described by non-linear parameter
combinations, the Fisher matrix approach tends to under-estimate the
error-bars.  Even with these limitations, the Fisher matrix approach
is invaluable to estimate degeneracies among parameters and assess
which data set combination can lift them.

%%%%%%%%%%%%%%%%%%%%%%%%%%%%%%%%%%%%%%%%%%%%%%%%%%%%%%%%%%%%%%%%%%%%%%%%%%%%%%%%
\subsection{Cluster model}
\label{subsec:model} 
%%%%%%%%%%%%%%%%%%%%%%%%%%%%%%%%%%%%%%%%%%%%%%%%%%%%%%%%%%%%%%%%%%%%%%%%%%%%%%%%

Galaxy clusters are the largest gravitationally bound structures in
the universe, and the properties of their dark matter halos should be
relatively insensitive to astrophysical processes, which typically
operate on scales much smaller than the size (i.e. virial radius) of a
massive cluster. However, processes such as radiative cooling, star
formation, heating and radiative feedback from active galactic nuclei,
turbulence, and non--thermal pressure support from energetic particles
accelerated in large--scale shocks, can all have significant impact on
the thermal state and spatial distribution of gas in the intra-cluster
medium (ICM), especially near the center of the cluster.  Many aspects
of the ICM remain poorly understood, despite extensive theoretical
work, numerical simulations, and high--resolution observations.  

There have been many approaches to building simplified models for
cluster structure. Some are purely phenomenological formulae for the
radial profiles, such as the simple 3--parameter ``beta--model''
(Cavaliere \& Fusco-Femiano 1976), or the 17--parameter generalized
NFW model proposed more recently by Vikhlinin et al. (2006), which
provides excellent fits to the range of observed X--ray profiles.
Many studies have based the models on physical ingredients, generally
assuming hydrostatic equilibrium, and parameterizing the the processes
listed above (see, e.g., Komatsu \& Seljak 2001; Voit et al. 2002;
Ostriker et al. 2005; Reid \& Spergel 2006; Fang \& Haiman 2008;
Ascasibar \& Diego 2008, and references therein).

In this paper, we do not attempt to build another {\it ab initio}
physical cluster model. Instead, we use a ``hybrid'' phenomenological
model, with physically motivated free parameters, similar to that
proposed in Reid and Spergel (2006) and Fang and Haiman (2008).  As we
will show below, this model can satisfy most available observational
constraints, and has the flexibility to include parameter variations.
The ICM properties are assumed to be spherically symmetric, on
average, and are determined by four factors: the radial entropy
profile, the profile of the gravitational potential, the equations of
hydrostatic equilibrium, and boundary conditions. Below, we describe
how we incorporate these four factors into our cluster model.

{\it Entropy profile.} The radial entropy profile is parameterized by
a power law,
\begin{eqnarray}
K(x)\equiv \frac{T}{\rho^{\gamma-1}}= \hat{K} K_{\mathrm{vir}} x^{s},
\label{eqn:entropy}
\end{eqnarray}
where $\rho$ and $T$ are density and temperature of the ICM gas,
$\gamma$ is the adiabatic index, which we choose to be 5/3,
appropriate for ideal monatomic gas, and $x$ is the dimensionless
radius, normalized by the virial radius $R_{\mathrm{vir}}$ of the
cluster. The virial radius $R_{\mathrm{vir}}$ is defined to be the
radius within which the mean density is equal to the virial density
$\rho_{\mathrm{vir}}$ determined from numerical simulations (see
equation \ref{eqn:rho_v} below).  $\hat{K}$ is the dimensionless
entropy at the virial radius, and $s$ quantifies the logarithmic slope
of the entropy with radius.  Note that convective stability requires
the entropy $K$ to be a monotonically increasing function of radius
(Voit et al. 2002), so we require $s\geq 0$.  The natural choice for
the entropy scale is its value estimated using virial theorem,
$K_{\mathrm{vir}}\equiv
T_{\mathrm{vir}}/(f_b\rho_{\mathrm{vir}})^{\gamma-1}$, where
$T_{\mathrm{vir}}=GM_{\mathrm{vir}}\mu m_p/(2R_{\mathrm{vir}})$.  In
above definitions, $f_b$ is the baryon fraction of the universe
($\Omega_b/\Omega_m$), $m_p$ is the mass of a proton, and $\mu$ is the
mean molecular weight of the ICM (we adopt 0.59 as its value,
appropriate for a fully ionized H-He plasma with helium mass fraction
equal to 0.25). Note that $K_{\mathrm{vir}}$ is the characteristic
entropy of a cluster in absence of non-gravitational forces rather
than entropy at virial radius; in particular, $\hat{K}>1$ even without
any feedback processes. Using the fitting formula given by Younger \&
Bryan (2007) and a slightly modified version of our current cluster
model code, we estimate that $\hat{K}$ is equal to 1.5 in the
self-similar case. 

{\it Dark matter halo gravitational potential.}  We assume that the
dark matter halos are spherically symmetric, and their density
profiles are described by the NFW shape. The assumption of spherical
symmetry can obviously be very inaccurate for individual clusters.  We
assume, however, that the main effect of the asymmetries is to
introduce a scatter in the global scaling relations, rather than to
change their mean (the accuracy of this assumption should be assessed
in the future in three--dimensional simulations of a large sample of
clusters).  The NFW profile is expressed as
\begin{eqnarray}
\rho_{\mathrm{DM}}(r)=\frac{\delta_{c}\rho_{c}}{(r/r_s)(1+r/r_s)^2}
\label{eqn:nfw}
\end{eqnarray}
where $r_s$ is the scale radius and $\rho_{c}$ is the critical density
of the universe. For a halo of DM mass $M_{\mathrm{DM}}$, the two
parameters $\delta_{c}$ and $r_s$ are determined from the
concentration parameter $c^{\mathrm{NFW}}$ and the virial density
$\rho_{\mathrm{vir}}$,
\begin{eqnarray}
r_s=\frac{R_{\mathrm{vir}}}{c^{\mathrm{NFW}}}=\frac{1}{c^{\mathrm{NFW}}}\left(\frac{3M_{\mathrm{DM}}}{4\pi \rho_{\mathrm{vir,DM}}}\right)^{1/3},\\
\delta_{c}=\frac{\rho_{\mathrm{vir,DM}}}{3\rho_{c}}\frac{(c^{\mathrm{NFW}})^3}{\mathrm{ln}(1+c^{\mathrm{NFW}})-c^{\mathrm{NFW}}/(1+c^{\mathrm{NFW}})}.
\label{eqn:r_delta}
\end{eqnarray}
Here $\rho_{\mathrm{vir,DM}}\equiv
(\Omega_{\mathrm{DM}}/\Omega_m)\rho_{\mathrm{vir}}$, with
$\Omega_{\mathrm{DM}}\equiv (\Omega_m-\Omega_b)$. We adopt a fitting
formula from the numerical spherical collapse
  model calculation by Kuhlen et al. (2005) for $\rho_{\mathrm{vir}}$,
\begin{eqnarray}
\rho_{\mathrm{vir}}=18\pi^2\Omega_{m}(z)\rho_{c}(z)[1+a\Theta^{b}(z)],
\label{eqn:rho_v}
\end{eqnarray}
where $\Theta(z)=\Omega_m^{-1}-1$, $a=0.432-2.001(|w(z)|^{0.234}-1)$,
and $b=0.929-0.222(|w(z)|^{0.727}-1)$, with $w(z)$ the dark energy
equation of state. Note that this formula differs slightly from other
expressions for the virial overdensity that are commonly used in the
literature (e.g., Bryan \& Norman 1998), and it includes an explicit
dependence on $w$. This feature is important to us, since we will
constrain $w$, and, as we will find below, this dependence drives the
constraints on $w(z)$ at low redshift. 

It is important to note that Equation~(\ref{eqn:rho_v}) was
obtained from spherical collapse calculations that assumed a constant
$w$. In particular, one may wonder whether this fitting formula is
still accurate when $w$ is redshift dependent. To check the goodness
of fit of Equation~(\ref{eqn:rho_v}), we performed numerical
calculations of the virial overdensity, using the spherical collapse
model described by Kuhlen et al.  We have found that
Equation~(\ref{eqn:rho_v}) is accurate to within 10\% for
time--varying $w$ models in the range of -1.5$<w_0<$-0.5, and $-\mid
0.5w_0\mid <w_a<\mid 0.5w_0\mid$.  More importantly, the virial
density computed with the numerical method is systematically more
sensitive to $w_a$ than Equation~(\ref{eqn:rho_v}) predicts.  For
example, we find that at redshifts $z_{\rm collapse}=0.1$, 0.2, and
0.4, the fractional change in $\rho_{\mathrm vir}(z_{\rm collapse})$,
when $w_a$ is changed from $w_a=0$ to $w_a=0.21$ (i.e., by its
$1\sigma$ value; see below), and all other parameters are held fixed,
is a factor of 3, 2, and 1.4 larger in the numerical calculation than
predicted by the fitting formula. The higher sensitivity is easy to
understand: $\rho_{\mathrm{vir}}$ at $z_{\rm collapse}$ depends on
$w(z>z_{\rm collapse})$, which at higher redshifts differs
increasingly from the constant value $w_0$. We therefore conclude that
using the Kuhlen et al. formula makes our constraints on $w_a$ below
conservative.

{\it Equations of hydrostatic equilibrium.} Below are the equations
that we solve to obtain the cluster gas density and pressure profiles
$\rho_g(r)$, $P(r)$. 
The first is from the hydrostatic equilibrium condition, the second is 
from mass conservation,
\begin{eqnarray}
\frac{dP(r)}{dr}=-\eta \rho_{g}(r)\frac{GM_{\mathrm{tot}}(<r)}{r^2},\\
\frac{dM_g(<r)}{dr}=4\pi r^2\rho_g(r),
\label{eqn:equilibrium}
\end{eqnarray}
where $M_{\mathrm{tot}}(<r)$ is the total
mass within radius $r$, including both dark matter and baryons,
and $M_g(<r)$ is the total gas mass
enclosed within radius $r$. The gas fraction $f_g$, which we will use
below, is defined to be
$M_g(<R_{\mathrm{vir}})/M_{\mathrm{tot}}(<R_{\mathrm{vir}})$. Finally,
the parameter $\eta\leq 1$ is introduced to allow for deviations from
strict hydrostatic equilibrium. Physically, deviations from $\eta=1$
could represent any non--thermal pressure support (e.g., from cosmic
rays and/or turbulence), and also lack of full virialization.  In
fact, allowing for turbulent support in the analytical model is known
to be necessary in order to reproduce the density and temperature
profiles for the ICM gas in simulations that include
non--gravitational pre--heating (Younger \& Bryan 2007).

{\it Boundary conditions.} The boundary condition at $r=0$ is
specified by requiring that $M_g(<r)$ is zero at the cluster
center. The boundary condition at the virial radius is imposed by
requiring that the gas pressure matches the momentum flux of the 
infalling gas,
\begin{eqnarray}
P(R_{\mathrm{vir}})=\frac{\Omega_b}{3(\Omega_m-\Omega_b)}b\rho_{\mathrm{DM}}(R_{\mathrm{vir}})v^{2}_{\mathrm{ff}}
\label{eqn:boundary}
\end{eqnarray}
where $v_{\mathrm{ff}}$ is the free--fall velocity from the turnaround
radius. Assuming the turnaround radius is twice the virial radius, as
in the spherical collapse model, we have
$v_{\mathrm{ff}}^{2}=GM_{\mathrm{vir}}/R_{\mathrm{vir}}$. We follow
Reid \& Spergel (2006) and introduce a free parameter $b$ to allow for
an uncertainty in this condition.

The cluster model described above has 5 free parameters that capture
uncertainties about cluster structure and evolution: $K$, $s$,
$c^{\mathrm{NFW}}$, $\eta$ and $b$.  All of these quantities could
additionally depend on both mass and redshift ($\eta$ and $s$ could
also explicitly depend on radius, which, however, we ignore here). We
use a power--law parameterization to allow for these dependencies,
\begin{eqnarray}
p=p_{\mathrm{norm}} \left(\frac{M}{M^{\star}}\right)^{p_{\mathrm{m}}}(1+z)^{p_{z}},
\label{eqn:parameter}
\end{eqnarray}
where $p$ could represent any of the 5 quantities. In equation
\ref{eqn:parameter}, each function is described by 3 parameters, one
for normalization, one for mass dependence and another for redshift
dependence. We choose $M^{\star}$ to be $10^{14} {\rm M_\odot}$ (this
choice is not essential, since changes to $M^{\star}$ can be
compensated by changes in the normalization).

Several cautionary remarks about the above model are in order.  First,
the assumption of power--law mass and $z$--dependence is likely valid
only when the variations over the observed mass and redshift range are
small; the real dependence could be more complicated, especially if a
wide mass or redshift range is considered. This could mean that actual
data will not be fit adequately by such power--laws; in this case,
additional parameters will likely have to be introduced (this
possibility is addressed more quantitatively below).

Second, unlike Reid \& Spergel (2006), we did not include additional
modeling of the cluster cores. The reason for this is that core
properties are known to vary significantly from cluster to cluster,
and it is difficult to capture this variation with a universal
parametrization.  Fortunately, neglecting the core makes relatively
little difference in our results. The two observables we focus on are
temperature $T$ and integrated Compton $y$ parameter $Y$. We checked
explicitly that introducing a flat entropy core within 0.1
$R_{\mathrm{vir}}$ changes the value of $Y$ by less than 2 percent
(this result is consistent with Reid \& Spergel 2006).  When computing
the emission--weighted temperature $T$, we excise the innermost
regions (see eq. \ref{eqn:tew} below), which makes our
temperature--observable similarly robust to core properties.  This cut
mimics the common procedures in existing observations, in which the
ambient gas temperature is inferred either by excising the core
region, or using a model (such as a cooling flow model) to eliminate
the contribution from core regions.  In order to minimize potential
biases from the model--dependence of such cuts, we use a simple
definition below.

Third, in reality, the cluster structure parameters will clearly
have cluster--to--cluster variations: each of the parameters appearing
in Equation~(\ref{eqn:parameter}) should therefore represent only a
mean value.  A scatter in any cluster structure parameter will induce
a scatter in the value of the observables we predict.  Below, we will
derive constraints only from the mean observables (i.e. our signal is
the mean $Y-T$ scaling relation; the finite distribution of $Y$ at
fixed $T$ is considered conservatively to be pure noise).  An
underlying scatter in a structure parameter can therefore have two
effects on our results. First, the measurement error of the mean
$\langle Y\rangle$ is increased, which will correspondingly weaken our
statistical constraints. Second, the mean inferred value of $\langle
Y\rangle$ can be biased, if the scatter in a parameter $p_i$
introduces a skewed $Y$-distribution -- and/or if the scatter is
large, and it introduces Malmquist bias (i.e. the low-$Y$ tail of the
clusters at fixed $T$ could be preferentially missing from the
sample).  The first of these effect will be addressed below by
allowing for a scatter in $Y$ itself; the possible biases from the
second effect are discussed in \S~\ref{subsec:results_3}.

In summary, our adopted baseline cluster model has $5\times 3$
parameters; given these parameters, we can numerically solve for the
density and pressure profiles, and deduce all other ICM quantities.
Below in \S~\ref{subsec:parameters}, we will use existing observational data and simulation results
to determine the fiducial values of these parameters.

%%%%%%%%%%%%%%%%%%%%%%%%%%%%%%%%%%%%%%%%%%%%%%%%%%%%%%%%%%%%%%%%%%%%%%%%%%%%%%%%
\subsection{Fisher matrix for the scaling relation}
\label{subsec:scaling}
%%%%%%%%%%%%%%%%%%%%%%%%%%%%%%%%%%%%%%%%%%%%%%%%%%%%%%%%%%%%%%%%%%%%%%%%%%%%%%%%

Our first way of constraining cosmological parameters is to use the
relation between the SZ flux $S_{\nu}$ (measured through the
integrated Compton $y$ parameter, $Y$) and the X--ray emission weighted
temperature $T_{\mathrm{ew}}$. The  electrons in the hot ICM gas scatter 
the cosmic microwave background (CMB) photons, which distorts the CMB spectrum
by the amount (e.g., Birkinshaw 1999; Carlstrom et al. 2002)
\begin{eqnarray}
S_{\nu}=j_{\nu}Y,
\label{eqn:szflux}
\end{eqnarray}
where $j_{\nu}$ is a known function of frequency,
\begin{eqnarray}
j_{\nu}(x)=2\frac{(k_{\mathrm{B}}T_{\mathrm{CMB}})^3}{(hc)^{2}}\frac{x^4e^x}{(e^x-1)^2}\left[\frac{x}{\mathrm{tanh}(x/2)}-4\right].
\label{eqn:jnu}
\end{eqnarray}
Here $x=h\nu/(k_{\mathrm{B}}T_{\mathrm{CMB}})$, $T_{\mathrm{CMB}}$ is
CMB temperature, $h$ is Planck constant, $k_{\mathrm{B}}$ is the
Boltzmann constant, and $c$ is the speed of light. $j_{\nu}$ is
positive at high frequency, negative at low frequency, and has a null
at $\nu\approx 220$ GHz. Physically, this means low--energy photons
are Compton scattered by the hot electrons to higher energies,
reducing the flux at low frequency and increasing it at high
frequency.

The ICM properties are all encoded in $Y$. The total distortion
within a fixed solid angle is given by
\begin{eqnarray}
Y(\le\theta)=2\pi\int_{0}^{\theta}y(\theta ')\theta ' d\theta ',
\label{eqn:yint}
\end{eqnarray}
where $y(\theta)$ is the Compton parameter along a given line of sight,
\begin{eqnarray}
y(\theta)=\frac{\sigma_{\mathrm{T}}}{m_ec^2}\int P_edl=\int\frac{\sigma_{\mathrm{T}}n_ek_{\mathrm{B}}T}{m_ec_{\star}^2}dl,
\label{eqn:y}
\end{eqnarray}
and $\sigma_{\mathrm{T}}$ is the Thompson cross section. In this work,
we use the value of $Y$ integrated over the whole cluster, i.e. in
equation~(\ref{eqn:yint}) we set $\theta=R_{\mathrm{vir}}/D_A$, with
$D_A$ the angular diameter distance. Combining equations
(\ref{eqn:yint}) and (\ref{eqn:y}), we get,
\begin{eqnarray}
Y=\frac{\sigma_{\mathrm{T}}k_{\mathrm{B}}}{m_ec^2D^2_A}\int n_eTdV_{\mathrm{cluster}}
\label{eqn:yint2}
\end{eqnarray}
where $V_{\mathrm{cluster}}$ is the volume of cluster. Equation
(\ref{eqn:yint2}) clearly shows that the integrated SZ flux directly
probes the thermal energy of ICM.

The emission--weighted temperature is calculated as
\begin{eqnarray}
  T_{\mathrm{ew}}=\frac{\int_{0.15R_{500}}^{R_{500}}4\pi r^2dr\int d\nu\rho^2_g(r)\Lambda(\nu,T)T(r)}{\int_{0.15R_{500}}^{R_{500}}4\pi r^2dr\int d\nu\rho^2_g(r)\Lambda(\nu,T)}
\label{eqn:tew}
\end{eqnarray}
where $\Lambda(\nu,T)$ is the cooling function, calculated by a
Raymond-Smith code (Raymond \& Smith 1977) with metallicity
$Z=0.3Z_{\odot}$, and $R_{500}$ is the radius with a mean enclosed
overdensity of 500 relative to the critical density. Note that we do
not integrate over the whole cluster -- instead, we excise the inner
region. This mimics the temperature measurements in X-ray
observations, which either excise the cooling flow regions, or model
and subtract their contribution. Since the inner radius (=$0.15
R_{500}$) has to be estimated from the data itself, this can introduce
uncertainties or biases in the inferred $T_{\mathrm{ew}}$. As we show
in \S~\ref{sec:discussion} below, in order not to degrade the
$1\sigma$ constraints we obtain below, the inner radius has to be
accurate (statistically) to within $\approx 5\%$, or the mass of the
cluster to within $\approx 15\%$.

Given a $Y(T)$ relation, 
we can construct the scaling--relation Fisher matrix for an
individual cluster as
\begin{eqnarray}
F^{\mathrm{sr.single}}_{ij}=\frac{1}{\sigma_{Y,T}^2}
\left.\frac{\partial Y}{p_i}\right|_{T_{\mathrm{ew}}}
\left.\frac{\partial Y}{p_j}\right|_{T_{\mathrm{ew}}} ,
\label{eqn:fsc_single}
\end{eqnarray}
where $p_i$ and $p_j$ are parameters to be constrained, $\sigma_{Y,T}$ is the total statistical uncertainty on the value
of $Y$, including both the intrinsic scatter $\sigma_i$ in $Y$ at
fixed $T$, and the measurement uncertainty in $Y$, $\sigma_m$,
$\sigma_{Y,T}^2=\sigma^2_{i}+\sigma^2_{m}$. Note that the partial
derivative is taken at a fixed temperature, not at given cluster mass,
since we are studying the relation of $Y$ {\it vs.}  $T_{\mathrm{ew}}$
not $Y$ {\it vs.} $M$.

For a sample of clusters for which $Y$ and $T$ are both measured, the
total Fisher matrix is the sum of the individual single--cluster
Fisher matrices.  We approximate this sum by an integration,
\begin{eqnarray}
F^{\mathrm{sr.total}}_{ij}=\Delta \Omega\int_{z_\mathrm{min}}^{z_\mathrm{max}}dz\frac{d^2V}{dzd\Omega}\int_{M_{min}(z)}^{\infty}dM \frac{dn}{dM}(M,z)F^{\mathrm{sr.single}}_{ij}
\label{eqn:fsc}
\end{eqnarray}
where $\Delta \Omega$, $z_{\mathrm{min}}$ and $z_{\mathrm{max}}$ are
the solid angle, and the minimum and maximum redshifts covered by the
survey, $M_{min}(z)$ is the mass of smallest detectable cluster at
each redshift, $d^2V/dzd\Omega$ is the comoving volume element, and
$dn/dM(M,z)$ is the halo mass function. The form of this Fisher matrix
is similar to the ``follow-up'' Fisher matrix used in Majumdar \& Mohr
(2003), except that their observable is the cluster mass itself (or a
mass--like quantity), whereas our observable here is $Y$. We used the
fitting formula by Jenkins et al. (2001) for the mass function (their
smoothed mass function, equation 9).  In the fitting formula in
Jenkins et al. (2001), the cluster mass $M$ is defined to be the mass
enclosed within a spherical region with overdensity of 180 with
respect to the mean background matter density, whereas we defined
clusters based on their virial overdensity with respect to the
critical density (eq.~\ref{eqn:rho_v} above). We used the NFW profile
to convert the Jenkins et al. mass function to be consistent with our
mass definition.

%%%%%%%%%%%%%%%%%%%%%%%%%%%%%%%%%%%%%%%%%%%%%%%%%%%%%%%%%%%%%%%%%%%%%%%%%%%%%%%%
\subsection{Fisher matrix for number counts}
\label{subsec:nc}
%%%%%%%%%%%%%%%%%%%%%%%%%%%%%%%%%%%%%%%%%%%%%%%%%%%%%%%%%%%%%%%%%%%%%%%%%%%%%%%%

Another way of constraining cosmological parameters is through the
cluster abundance. The observable in this case is the number of
clusters in a given range of redshift and $Y$,
\begin{eqnarray}
N_{i\alpha}=\Delta \Omega \Delta z\frac{d^2V}{dzd\Omega}(z_i)\int_{M_{\mathrm{min}}}^{\infty}g(Y_{\alpha},M)\frac{dn(M, z_i)}{dM}dM
\label{eqn:nc}
\end{eqnarray}
where $M_{\mathrm{min}}$ is a minimum mass we impose by hand
(representing a sharp survey selection threshold; see discussion in
\S~\ref{subsec:results_3} below for allowing uncertainties in the
selection), and $g(Y_{\alpha}, M)$ is the probability that a cluster
with mass $M$ has a value of $Y$ within the range of the $\alpha$th
$Y$--bin. In this paper, for simplicity, we assume Gaussian scatter
between $M$ and $Y$, so that $g(Y_{\alpha}, M)$ has an analytical
form. Suppose bin $Y_{\alpha}$ is specified by its minimum
$Y_{\mathrm{min}}$ and maximum $Y_{\mathrm{max}}$, and for a given
mass $M$, $Y$ has a mean $\bar Y(M)$ and r.m.s. of $\sigma_{Y,M}$.  In
this case,
\begin{eqnarray}
\label{eqn:g}
g(Y_{\alpha}, M)&=&\int_{Y_{\mathrm{min}}}^{Y_{\mathrm{max}}}\frac{1}{\sigma_{Y,M}\sqrt{2\pi}}\exp\left\{-\frac{[Y-Y(M)]^2}{2\sigma_{Y,M}^2}\right\}dY \\\nonumber
&=&\frac{1}{2}\left\{{\rm erf}\left[\frac{Y_{\mathrm{max}}-Y(M)}{\sigma_{Y,M}\sqrt{2}}\right]-{\rm erf}\left[\frac{Y_{\mathrm{min}}-Y(M)}{\sigma_{Y,M}\sqrt{2}}\right]\right\}
\end{eqnarray}
Assuming Poisson errors dominate in the number counts
($\sigma_{N}=\sqrt{N}$), and summing over all redshift-- and
$Y$--bins, the total Fisher matrix for the cluster abundance (Holder
et al. 2001) is given by
\begin{eqnarray}
F_{ij}^{nc.total}=\sum_{\alpha=1}^{N_z}\sum_{\beta=1}^{N_y}\frac{1}{N_{\alpha\beta}}\frac{\partial N_{\alpha\beta}}{\partial p_i}\frac{\partial N_{\alpha\beta}}{\partial p_j},
\label{eq:fnc}
\end{eqnarray}
where $N_z$ and $N_y$ are the number of redshift bins and $Y$ bins,
respectively. This expression ignores sample variance, whose effect on
the cluster abundance constraints has been considered in detail in
previous works (Hu \& Kravtsov 2003; Lima \& Hu 2004; Fang \& Haiman
2006), and has been found to be modest, especially if the survey is
sub--divided into many angular cells, and the variance is considered
as signal, rather than noise; Lima \& Hu 2004).  Likewise, Holder et
al. (2001) explored the validity of the Fisher matrix approach for
forecasting cluster count constraints, and found it to be a good
approximation, with the exception of the constraints for $\sigma_8$.

%%%%%%%%%%%%%%%%%%%%%%%%%%%%%%%%%%%%%%%%%%%%%%%%%%%%%%%%%%%%%%%%%%%%%%%%%%%%%%%%
\subsection{Fiducial parameter  values}
\label{subsec:parameters}
%%%%%%%%%%%%%%%%%%%%%%%%%%%%%%%%%%%%%%%%%%%%%%%%%%%%%%%%%%%%%%%%%%%%%%%%%%%%%%%%

Here we summarize all the parameters in this work, and explain our
choice of their fiducial values. Overall, the model parameters can be
grouped into three categories: cosmological parameters, cluster model
parameters and survey parameters.

{\it Cosmological parameters.} We include the following 8 standard
cosmological parameters, with the 3--year results from the {\it
  Wilkinson Microwave Anisotropy Probe (WMAP)} experiment as their
fiducial values (Spergel et al. 2007): $\Omega_m=0.244$, $\Omega_{DE}=0.756$,
$\Omega_{b}=0.0413$, $h=0.72$, $w_0=-1$, $w_a=0$, $\sigma_8=0.76$ and
$n_s=0.96$. Here $w_0$ and $w_a$ parametrize the dark energy equation
of state,
\begin{eqnarray}
w(z)=w_0+w_a(1-a)=w_0+w_a\frac{z}{1+z}.
\end{eqnarray}
We do not assume a spatially flat universe, so the 8 cosmological
parameters are independent.  The more recent 5--year results from {\it
  WMAP} are consistent with the 3--year results. $\sigma_8$ is
slightly higher ($\sigma_8=0.817$ from the combination of the {\it
  WMAP} result with baryon acoustic oscillations and supernovae;
Dunkley et al. 2008). Adopting this new value would increase the
number of detectable clusters, and tighten our constraints below.  We
emphasize that the number counts constrain all 8 parameters directly,
while the scaling relation alone can only constrain 6 of them
($\sigma_8$ and $n_s$ do not affect scaling relation). Nevertheless,
when combined with the number counts, the information from the scaling
relation can indirectly help constrain $\sigma_8$ and $n_s$, by
breaking degeneracies, as we will demonstrate in \S~\ref{sec:results}
below.

{\it Cluster model parameters.} The cluster model described in \S~
\ref{subsec:model} above has $5\times3=15$ parameters, describing the
normalization $\hat{K}$ and logarithmic slope $s$ of the gas entropy
profile, the concentration $c^{\mathrm{NFW}}$ parameter for the dark
matter halo profile, any contributions from non--thermal pressure
$\eta$, and the gas pressure at the virial radius $b$, each with a
normalization, redshift dependence and mass dependence. We set their
fiducial values using results from simulations and observations.

The self-similar collapse model that invokes only gravity and shock
heating predicts a universal entropy profile $K(r)\propto r^{1.1}$
(Tozzi \& Norman 2001; Borgani et al. 2001), which is in general
agreement with observations outside the core (Ponman et al. 2003;
Pratt et al. 2006). We therefore adopt this fiducial value for $s$,
with no dependence on mass or redshift
($s_{\mathrm{m}}=s_{\mathrm{z}}=0,s_{\mathrm{norm}}=1.1$).

The difference in the cluster mass inferred from weak lensing and
X-ray measurements suggests that non--thermal pressure contributes
about 10\% to total gas pressure (Zhang et al. 2008), similar amounts
have also been seen in simulations (Rasia et al. 2004; Kay et
al. 2004; Faltenbacher et al. 2005), thus we adopt $\eta=0.9$ with no
mass and redshift dependence.

For the concentration parameter $c^{\mathrm{NFW}}$, we adopt the
fiducial value that is directly computed from cosmological simulations
in Voit et al. (2003),
\begin{eqnarray}
c^{\mathrm{NFW}}(M,z)=8.5\left(\frac{M}{10^{15}h^{-1}{\rm M_{\odot}}}\right)^{-0.086}(1+z)^{-0.65}.
\label{eqn:concentration}
\end{eqnarray}
Note that unlike our other model parameters, in principle,
$c^{\mathrm{NFW}}$ can be accurately computed ab--initio, using
three--dimensional N--body or hydro simulations.  However, the
uncertainties are still significant, and the current results are in
tension with X--ray observations (e.g., Duffy et al. 2008); for
completeness, we therefore include it as a free parameter in our baseline
model.  Below, we will investigate the benefits of placing tight
priors on this parameter (and we find the benefits to be small).  For
simplicity, we set $b=1$, which corresponds to the condition that all
kinetic energy is transformed into thermal energy at the virial
radius.  Molnar et al. (2008) recently studied in detail the
morphology and properties of virial shocks around galaxy clusters in
smooth particle hydrodynamics (SPH) and adaptive mesh refinement (AMR)
simulations of a sample of individual clusters. Although virial shocks
are often preceded by external shocks farther out, closer to $2-3
R_{\mathrm vir}$, a significant fraction of the clusters' surface area
is covered by strong virial shocks located at $\sim R_{\mathrm vir}$,
for which $b=1$ should be a good approximation. While detailed
simulations could help refine the best fiducial choice for the mean
boundary pressure, we do not expect the choice of this fiducial value
to have a significant impact on our forecasts.

\begin{figure}
\begin{tabular}{c}

\rotatebox{-0}{\resizebox{90mm}{!}{\includegraphics{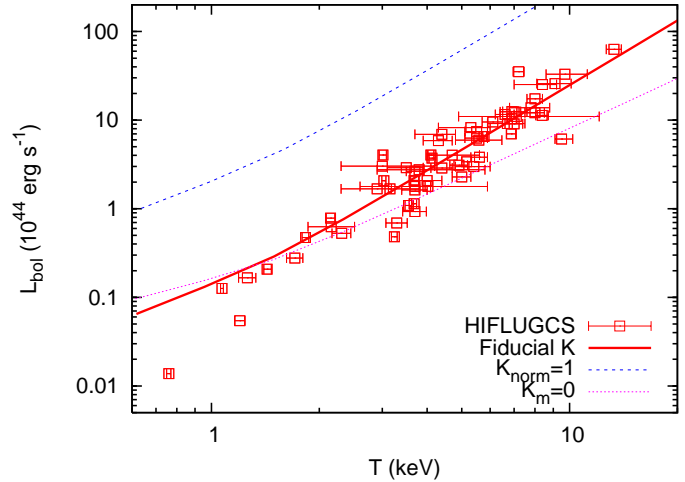}}}
\end{tabular}
\caption{The $L-T$ relation observed at low redshift, together with
  the predictions in our cluster model. The data points are from the
  HIFLUGCS cluster sample, with average redshift $z=0.05$ (Reiprich \&
  B\"ohringer 2002). The thick solid [red] curve shows the $L-T$
  relation at this redshift, predicted in our fiducial cluster model
  with $\hat{K}_{\mathrm{norm}}=2.4$, $\hat{K}_{\mathrm{m}}=-0.12$ and
  $\hat{K}_{\mathrm{z}}=0.0$. For comparison, we also show $L-T$ curve for
  $\hat{K}_{\mathrm{norm}}=1$ (dashed [blue] curve, $\hat{K}_{\mathrm{m}}$ and $\hat{K}_{\mathrm{z}}$ are kept at their
fiducial values) and for
  $\hat{K}_{\mathrm{m}}=0.0$ (dotted [thin red] curve, $\hat{K}_{\mathrm{norm}}$ and $\hat{K}_{\mathrm{z}}$ are kept at their
fiducial values). Decreasing
  $\hat{K}_{\mathrm{norm}}$ increases the luminosity at a given temperature,
  whereas changing $\hat{K}_{\mathrm{m}}$ changes the slope. }
\label{fig:lt}
\end{figure}

\begin{figure}
\begin{tabular}{c}

\rotatebox{-0}{\resizebox{90mm}{!}{\includegraphics{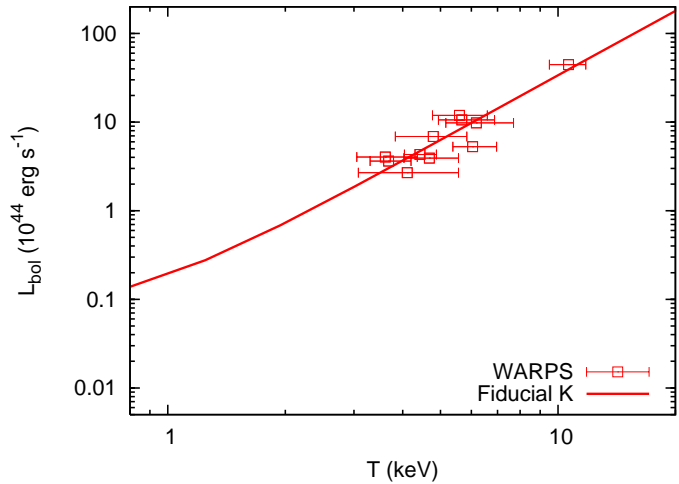}}}
\end{tabular}
\caption{The $L-T$ relation observed at high redshift ($z\approx
  0.8$), and predictions in our cluster model. The data--points are
  from the high--redshift WARPS clusters with average redshift
  $\langle z \rangle = 0.8$. The thick solid [red] curve is the
  prediction in our fiducial cluster model with
  $\hat{K}_{\mathrm{norm}}=2.4$, $\hat{K}_{\mathrm{m}}=-0.12$ and
  $\hat{K}_{\mathrm{z}}=0.0$.}
\label{fig:lth}
\end{figure}

The fiducial value of $\hat{K}$ is fixed by fitting the $L - T$
relation. In Figure \ref{fig:lt}, we compare the observed $L-T$
relation to the predicted values for the best-fit $\hat{K}$. The data
points are from the HIFLUGCS cluster sample (Reiprich \& B\"ohringer
2002). The bolometric luminosity is computed from
\begin{eqnarray}
L=\int dV\int d\nu n_e(r)n_H(r) \Lambda(T,\nu),
\end{eqnarray}
where $n_e$ and $n_H$ are electron and proton number densities,
respectively, and we assume a helium mass fraction of 25\%. Unlike
$Y$, $L$ is sensitive to the core properties, and the HIFLUGCS sample
includes both cooling--core and non--cooling--core clusters. To
account for this mixing, we modify the entropy profile in our model
clusters, and add a flat entropy core of size $0.1R_{\mathrm{vir}}$.
The best--fit value of $\hat{K}$ is found, by minimizing $\chi^2$
using the data from Reiprich \& B\"ohringer, to be
$2.4(M/M^{\star})^{-0.12}(1+z)^0$.

This is in accordance with previous results on cluster formation and
evolution. In particular, we find that the entropy is elevated
($\hat{K}_{\mathrm{norm}}>1.5$), as expected by feedback processes,
and also that the increase in entropy is more significant for
lower--mass clusters, breaking the self similar relation. For
comparison, in the same figure we also plot the $L-T$ curves expected
for $\hat{K}_{\mathrm{norm}}=1.0$ (dashed line, 
$\hat{K}_{\mathrm{m}}$ and $\hat{K}_{\mathrm{z}}$ are kept at their 
fiducial values) and
$\hat{K}_{\mathrm{m}}=0$ (dotted line, $\hat{K}_{\mathrm{norm}}$ and $\hat{K}_{\mathrm{z}}$ are kept at their fiducial values).
Lowering $\hat{K}_{\mathrm{norm}}$ raises $L$ at a given temperature,
while lowering $\hat{K}_{\mathrm{m}}$ increases the slope.

We emphasize that $\hat{K}=1$, as mentioned above, does not correspond
to the gravitational-heating only case, and that, in agreement with
previous results, the most massive clusters in Figure \ref{fig:lt}
fall on the observed $L-T$ relation even without any non-gravitational
heating.  To verify this, we followed Fang \& Haiman (2008), and used
the fitting formula given in Younger \& Bryan (2007) to compute the
maximum entropy in the gravitational-heating only case. This entropy is
found to be $\approx 1.5K_{\rm vir}$ at the virial radius. With our
adopted set of cosmological parameters, a 10keV cluster has mass of
$3\times 10^{15}~{\rm M_\odot}$, and $\hat{K}=1.6$; the difference
between our fiducial $K$ and that of the simulation is only 0.1 around
$T=10$ keV (this difference is due to the fact that our fiducial
entropy profile is steeper than found in adiabatic simulations).

Though a successful fit in general, at the lowest temperatures shown
in Figure~\ref{fig:lth} ($T\lsim 1$keV), the best--fit $L-T$ relation
has a slope that flattens slightly, in contrast with observations that
indicate a steepening at these temperatures (e.g., Helsdon \& Ponman
2000). This shows that our particular power--law parameterization
(eq.~\ref{eqn:parameter}) is insufficient in capturing the increase in
the mean entropy at low temperatures.  Correcting this deficiency
would be possible by using different parameterizations (e.g.,
parameterizations that preferentially affect the cores of
low--temperature clusters, such as the ``entropy--floor'' models; e.g.,
Fang \& Haiman 2008). However, in this paper, we focus on clusters
detectable in SZ experiments. These have a mass $\gsim 10^{14}h^{-1} {\rm
M_{\odot}}$ (or equivalently, a mean temperature of $T\gsim 1.5keV$),
and the power--law models provide a good fit the $L-T$ relation of
these clusters (Fig.~\ref{fig:lth}).  

We also note that we do not find a need for $\hat{K}$, as expressed in the
form of equation~(\ref{eqn:parameter}), to evolve with redshift.  In
Figure~\ref{fig:lth}, we show the $L-T$ relation predicted in our
model, assuming $\hat{K}_{\mathrm{z}}=0$, at redshift $z=0.8$, together with
data from the high--redshift cluster sample from the Wide Angle ROSAT
Pointed Survey (WARPS; with average redshift of $\langle z\rangle
=0.8$). As the figure shows, the model provides an excellent match to
the data. Although not immediately obvious, this conclusion is
qualitatively in agreement with the result in Fang \& Haiman (2008),
who found that that if a fixed entropy floor is assumed to exist in
all cluster at a given redshift, then this floor value has to decrease
toward higher redshifts. The entropy floor is the difference between
total entropy and baseline value (without non--gravitational
heating). Our result indicates that the ratio of total entropy to
baseline entropy is the same for low and high redshifts. This means that
the difference is smaller for high--redshift clusters, since they have
a higher density and a smaller baseline entropy.  (For more details on
this apparent coincidence, see Figure 6 and the related discussion in
Fang \& Haiman 2008).

\begin{figure}
\begin{tabular}{c}

\rotatebox{-0}{\resizebox{90mm}{!}{\includegraphics{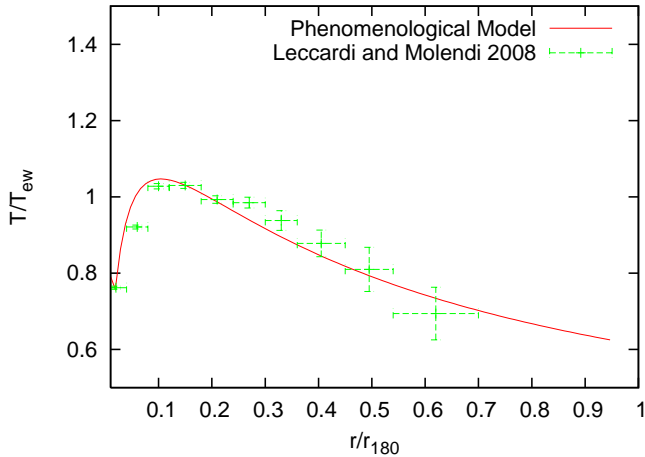}}}
\end{tabular}
\caption{Radial temperature profile in our fiducial phenomenological
model, in which the entropy profile has a logarithmic slope of
$s=1.1$, compared to recent XMM-Newton measurements of Leccardi \&
Molendi (2008).  We adopt a redshift of $z=0.2$, approximately the
median redshift of the observed cluster sample, and a mass of
$M=10^{15}{\rm M_{\odot}}$, which corresponds to a temperature of
approximately 6 keV, close to the temperature of a typical cluster in
the sample.}
\label{fig:tprofile}
\end{figure}

Our parameterization allows us to vary the slope of the entropy
profile $s$, which is necessary to fit temperature profiles measured
in X--ray observations. This, in fact, is the main advantage of our
parameterization compared to similar models that include a constant
entropy floor, since the latter approach generically fails to match
radial profiles (e.g., Younger \& Bryan 2007 and references therein).
In Figure~\ref{fig:tprofile}, we show the temperature profile in our
fiducial cluster model with $s=1.1$, compared to the mean profile
recently inferred from XMM-Newton observations (Leccardi \& Molendi
2008). In this plot, we adopt $z=0.2$, which is approximately the
median redshift of the XMM-Newton cluster sample, and $M=10^{15}{\rm
M_{\odot}}$, which corresponds to a mean temperature about 6 keV,
approximately the temperature of a typical cluster in the sample. We
follow Leccardi \& Molendi (2008) to compute $R_{180}$ as
\begin{eqnarray}
R_{180}=1780\left(\frac{T}{5\mbox{keV}}\right)^{1/2}h(z)^{-1}\mbox{kpc}.
\label{eqn:r_180}
\end{eqnarray}
As shown in Figure \ref{fig:tprofile}, the model temperature profile
is in good agreement with observations out to the radius of $0.6
R_{180}$ where data is available.

\begin{figure}
\begin{tabular}{c}

\rotatebox{-0}{\resizebox{90mm}{!}{\includegraphics{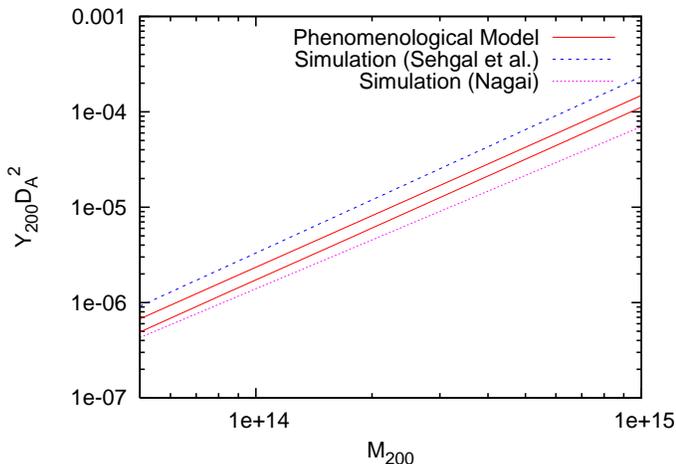}}}
\end{tabular}
\caption{Scaling relation of $Y_{200}D_A^2-M_{200}$ of our phenomenological model (Solid line) and two simulations--Sehgal et al.(2007, dashed line) and Nagai (2006, dotted line). The upper solid line is $Y_{200}$ over a cylindrical region in the same way as computed in Sehgal et al. (2007), while the lower line is that over a sphere in the same way as computed in Nagai 2006. The redshift is set to be 0 here.}
\label{fig:y_m}
\end{figure}

Finally, we compare the $Y-M$ relation in our fiducial model with
predictions from simulations.  More specifically, in
Figure~\ref{fig:y_m}, we show the $Y_{200}D_A^2-M_{200}$ relation in
our fiducial model, together with the predictions for the same
quantity in Nagai (2006) and Sehgal et al. (2007), who respectively
use high--resolution hydrodynamical simulations, and N-body
simulations of dark matter halos and a prescription for the
corresponding gas distribution.  Here $Y_{200}$ and $M_{200}$ are the
SZ Compton $y$ parameter and the total mass within the radius
$R_{200}$. We note that Sehgal et al. integrate over a cylindrical
region extending to an angular radius corresponding to $R_{200}$,
while Nagai integrates over a sphere of radius $R_{200}$. For a fair
comparison, we compute $Y_{200}$ both ways. The upper solid [red]
curve in Figure~\ref{fig:y_m} is $Y_{200}$ in a cylindrical region,
while the lower solid [red] curve is that in a sphere.  The redshift
is set to be $z=0$, to match Nagai's simulation.  We use the fitting
parameters in row 2 of Table 2 in Sehgal et al. (2007), since we focus
on clusters above the SZ detection threshold (about
$3\times10^{14}{\rm M_{\odot}}$ at intermediate redshifts). Sehgal et
al. find a slightly steeper slope than Nagai, which they attribute to
the effect of AGN heating. The slope of Nagai roughly agrees with the
self-similar expectation. Our slope is closer to that of Sehgal et
al., which includes star formation, and AGN feedback, but no
cooling. Overall, the slopes, however, are close to one another, and
our normalization falls between the predictions of the two
simulations.

In addition to the 15 parameters fixed above, we also allow for
independent scatter in the $Y-T$ and the $Y-M$ relations.  We chose a
fiducial value of 10\% for both (i.e. $\sigma_i$ in
eq.~\ref{eqn:fsc_single} and $\sigma_{Y,M}$ in eq.~\ref{eqn:g} are both
$0.1Y$), motivated by the simulations of Nagai (2006), who finds an
r.m.s. scatter between $Y$ and $M$ of 10-15\%.  The effect of a
non--zero $Y-M$ scatter on our results is two--fold. Scatter increases
the number of detected cluster at a given flux threshold (because of
the steep slope of the mass function, more clusters scatter from below
the threshold to above it than vice-versa), which is helpful in
constraining cosmology.  On the other hand, scatter flattens the
effective mass function, which degrades the information derivable from
the shape of the mass function (as will be demonstrated in
\S~\ref{sec:results} below, we find that the second effect dominates
in the constraints from the cluster counts).

In summary, we conclude that our fiducial cluster structure model
matches existing observations and simulations reasonably well, at
least at low redshifts, and for the clusters above the expected
detection threshold of future SZ surveys.  This gives us confidence to
use our adopted model parameterization to forecast cosmological
constraints, and to study the effect of cluster structure
uncertainties.  Of course, in the future, as better data becomes
available, it is possible (indeed likely) that modifying the
parameterization of the cluster structure model will become necessary.

{\it Survey parameters.}  Survey parameters include the following: sky
coverage $\Delta \Omega$, frequency $f$, measurement noise $\sigma_m$,
redshift range $z_{\mathrm{min}}-z_{\mathrm{max}}$ and related
parameters, such as redshift bin size and $Y$ bin size. We adopt
typical values relevant to upcoming SZ surveys for these parameters.

The sky coverage $\Delta \Omega$ is set to be 4,000 $\mathrm{deg}^2$
which is the solid angle covered by SPT in 2 years.
The frequency $f$ is chosen to be $145 GHz$, where the SZ signal
reaches maximum decrement. Most surveys will observe in bands at
multiple frequencies, in order to separate the SZ signal from other
CMB secondary anisotropies and from foregrounds. However, almost all
planned surveys have a frequency band around $145 GHz$.
The detector noise $\sigma_m$ is set to be $1 mJy$, which represents a
typical value for the total SZ flux of the smallest detectable cluster
in upcoming surveys (such as in SPT). At frequency of $145 GHz$, this
corresponds to $\sigma_m=9.36\times10^{-13}$sr. We set the cluster
detection threshold to be $5\sigma_m$. At low redshift, surface
brightness becomes an important additional detection criterion, and a
simple flux limit becomes inadequate, so we impose a floor on the mass
limit $M_{\mathrm{min}}=10^{14}h^{-1} {\rm M_{\odot}}$
(eq.~\ref{eqn:nc}). We will discuss how this choice affect our result
in \S~\ref{sec:discussion}.

We further assume the SZ survey covers the redshift range $0.0<z<2.0$,
and we divide this range into 40 uniform bins ($\Delta z=0.05$). This
of course requires the cluster redshift measurement uncertainty is
better than 0.05. This accuracy should be achievable by follow--up
surveys designed for this purpose; for example, the Dark Energy Survey
can determine cluster photometric redshifts to an accuracy of 0.02 or
better out to $z\sim 1.3$ (Abbott et al. 2005). 

Finally, we divide the $Y$ range into 8 bins, which we allocate so
that each bin contains a similar number of clusters.  This requirement
led us to adopt the following boundaries of the $Y$--bins, in units of
the $5\sigma_m$ detection threshold:
      $[1-2^{1/4}]$,  
$[2^{1/4}-2^{1/2}]$,  
$[2^{1/2}-2^{3/4}]$,  
$[2^{3/4}-2^{1.0}]$,  
$[2^{1.0}-2^{5/4}]$,  
$[2^{5/4}-2^{3/2}]$,  
$[2^{3/2}-2^{7/2}]$,   
$[2^{7/4}-\infty]$.

\begin{figure}
\begin{tabular}{c}

\rotatebox{-0}{\resizebox{90mm}{!}{\includegraphics{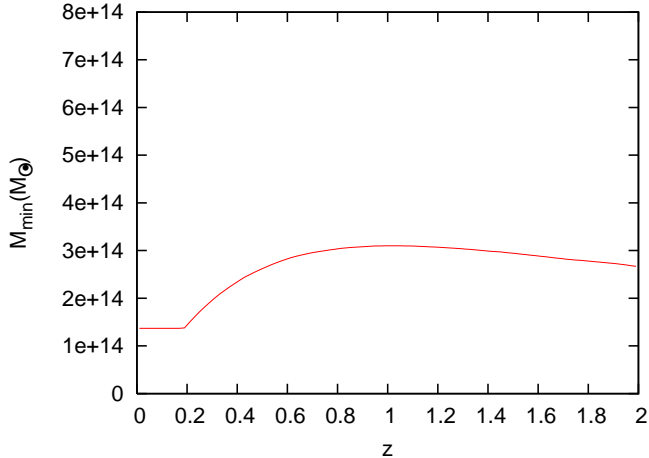}}}
\end{tabular}
\caption{The mass of the smallest detectable cluster as function of
redshift (neglecting scatter), corresponding to a constant SZ flux of
5 mJy. The flat part at low redshift is an additional constant mass
limit imposed by hand, $10^{14}h^{-1} {\rm M_{\odot}}$.}
\label{fig:m_min}
\end{figure}

In Figure \ref{fig:m_min}, we show the minimum mass detectable by the
SZ survey as function of redshift, neglecting scatter between $Y$ and
$M$. The flat portion of the curve at low redshift is the limit
imposed by hand ($10^{14}h^{-1} {\rm M_{\odot}}$). The total number of
detected clusters is about 6,800 for our adopted set of fiducial
cosmological parameters (for reference, we note that the total number
would be a factor of $\sim 3$ higher, $\approx 20,000$, if we instead
adopted the 1st year {\sl WMAP} cosmological parameters; the
difference attributable mostly to the change in $\sigma_8$ from 0.90
to 0.76).

In our fiducial calculation, we assume that temperature measurements
are available for all clusters detected by the SZ survey.  This is
usually taken to require X--ray spectroscopic data for each cluster.
Mroczkowski et al. (2008) find that a joint analysis of SZ and X-ray
imaging data, for 3 clusters detected with the SZA instrument, yields
temperature profiles that are in good agreement with spectroscopic
X-ray measurements.  This may ease the requirement on the depth of the
X--ray survey. The availability of temperatures for a large fraction
of the SZ clusters may, however, still be an optimistic assumption,
since the X--ray flux drops much faster than SZ flux at high
redshift. We discuss the effect of partial followup in
\S~\ref{sec:discussion}.

%%%%%%%%%%%%%%%%%%%%%%%%%%%%%%%%%%%%%%%%%%%%%%%%%%%%%%%%%%%%%%%%%%%%%%%%%%%%%%%%
\section{Results}
\label{sec:results}
%%%%%%%%%%%%%%%%%%%%%%%%%%%%%%%%%%%%%%%%%%%%%%%%%%%%%%%%%%%%%%%%%%%%%%%%%%%%%%%%

With the cluster model and the Fisher matrix technique described
above, we are now ready to forecast constraints from upcoming SZ and
X--ray surveys.  As an academic exercise, in \S~\ref{subsec:results_c}
we first consider an ``idealized case'', in which cluster parameters
are precisely known, and only cosmological parameters are constrained.
This exercise serves two purposes: (i) it allows us to understand
where the cosmology sensitivity comes from, and (ii) it will clarify
the amount of the degradation in the constraints, once the cluster
model parameter uncertainties are included.  In
\S~\ref{subsec:results_2}, we relax the assumption that cluster
structure parameters are known, and simultaneously constrain
cosmological and cluster parameters.  Finally, in
\S~\ref{subsec:results_3}, we investigate the effects of the
additional uncertainties in the scatter and incompleteness.

%%%%%%%%%%%%%%%%%%%%%%%%%%%%%%%%%%%%%%%%%%%%%%%%%%%%%%%%%%%%%%%%%%%%%%%%%%%%%%%%
\subsection{Constraints with cosmological parameters alone}
\label{subsec:results_c}
%%%%%%%%%%%%%%%%%%%%%%%%%%%%%%%%%%%%%%%%%%%%%%%%%%%%%%%%%%%%%%%%%%%%%%%%%%%%%%%%

Table~\ref{tbl:constraints1} lists the marginalized 1$\sigma$ errors
on cosmological parameters. In this and in the other tables below,
``SR'' stands for ``scaling relations'', and ``NC'' stands for
``number counts''.  In addition to the marginalized errors (computed
as $\sigma=\sqrt{(F^{-1})_{ii}}$), in Table~\ref{tbl:constraints1} we
also list the single--parameter errors $(F_{ii})^{-1/2}$ , and the
degeneracy parameter $\cal{D}$, which we define as
$\sigma/(F_{ii})^{-1/2}$.  In the limit of no degeneracies,
$\cal{D}\rightarrow$ 1, while large $\cal{D}$ indicates significant
degeneracy.

\begin{table*}
  \caption{Estimated 
1$\sigma$ errors on the cosmological
parameters in the academic case when cluster structure parameters are
precisely known. Here ``SR'' and ``NC'' stand for ``scaling relation''
and ``number counts'', respectively. The degeneracy parameter
$\cal{D}$ is defined as $\sigma/(F_{ii})^{-1/2}$, so that large values
indicate significant degeneracy.
}
    \label{tbl:constraints1}
\begin{center}
\begin{tabular}{ c c c c c c c c c c c}
 \hline\hline
Parameter constraints & \multicolumn{3}{c}{SR} & \multicolumn{3}{c}{NC}& \multicolumn{4}{c}{Combined}\\
 & $\sigma$& $(F_{ii})^{-1/2}$ & $\cal{D}$& $\sigma$& $(F_{ii})^{-1/2}$ & $\cal{D}$&$\sigma$ &$(F_{ii})^{-1/2}$ & $\cal{D}$&$\xi^{\dagger}$\\
\hline
$ \Omega_m$ & 0.055 & 0.00030 & 183.4 & 0.023 & 0.0024 & 9.6&0.009&0.00029&29.6&0.17\\
$ \Omega_{DE}$ & 0.20 & 0.028 & 7.0 & 0.29 & 0.016 & 18.4 &0.06&0.014&4.32&0.13\\
$ \Omega_b$ & 0.012 & 0.00006 & 189.7 & 0.007 & 0.00035 & 20.2 &0.0037&0.00006&60.4&0.36\\
$ h$ & 0.08 & 0.0014 & 55.1 & 0.11 & 0.006 & 18.6 &0.050&0.0014&36.4&0.62\\
$ w_0$ & 0.037 & 0.010 & 3.6 & 0.20 & 0.018 & 11.1 &0.016&0.009&1.8&0.20\\  
$ w_a$ & 0.21 & 0.044 & 4.7 & 1.4 & 0.10 & 14.4 & 0.11&0.040&2.7&0.29\\
$ \sigma_8$ & N/A & N/A & N/A & 0.016 & 0.0011 & 14.2 &0.007&0.0011&6.2&0.19\\
$ n_s$ & N/A & N/A & N/A &0.13 & 0.023 & 5.7  & 0.036&0.022&1.6&0.08\\
 \hline
\end{tabular}

\raggedright $^{\dagger}$ ``Complementarity'' parameter which quantifies the level
of degeneracy breaking when different measurements are combined. 
See eq. \ref{eqn:comparameter} and Fang \& Haiman (2008) for the formal definition. 
for details.
\end{center}
\end{table*}

As Table \ref{tbl:constraints1} shows, the scaling relation in general
has a constraining power comparable to the number counts.  For some of
the cosmological parameters, and especially for $w_0$ and $w_a$, the
SR is even more powerful than the NC approach. This might be
surprising, since the cluster abundance is known to be exponentially
sensitive to cosmological parameters, while the SR depends on these
parameters more--or--less ``linearly''.  However, the scaling relation
approach has its own advantages. We compare these two approaches in
more detail in \S~\ref{sec:discussion}. Let us first see where the
constraints come from in the scaling relation approach.

It is easy to see from equation~(\ref{eqn:yint2}) above that
\begin{eqnarray}
Y\propto D_A^{-2}M_{g}T \sim D_A^{-2}f_g M_{\mathrm{vir}}T.
\label{eqn:explain1}
\end{eqnarray}
Since we are studying the $Y-T$ relation, we eliminate the mass
$M_{\mathrm{vir}}$ from this equation by converting $M_{\mathrm{vir}}$ to
$T$ and $\rho_{\mathrm{vir}}$ using the virial theorem and mass
conservation,
\begin{eqnarray}
T_{\mathrm{vir}}\propto M_{\mathrm{vir}}/R_{\mathrm{vir}}
\label{eqn:explain2_1}\\
M_{\mathrm{vir}}\propto \rho_{\mathrm{vir}}R^3_{\mathrm{vir}}.
\label{eqn:explain2_2}
\end{eqnarray}
Combining the above three equations, we find
\begin{eqnarray}
Y\propto  D_A^{-2}f_g \rho_{\mathrm{vir}}^{-1/2} T^{5/2}.
\label{eqn:explain3}
\end{eqnarray}
Equation~(\ref{eqn:explain3}) indicates that the dependence on
cosmological parameters can arise through three terms: the angular
diameter distance $D_A$, the gas fraction $f_g$ (defined in
\S~\ref{subsec:model} above), and the virial overdensity
$\rho_{\mathrm{vir}}$.  Here $f_g$ and $\rho_{\mathrm{vir}}$ are both
related to cluster properties; $D_A$, on the other hand, is a direct
property of space--time geometry. Below, we first study the dependence
of $Y-T$ through the grouped combination of $f_g
\rho_{\mathrm{vir}}^{-1/2}$ and through $D_A^{-2}$.  
This grouping is useful because
$D_A$ is a pure geometrical quantity, and the cosmology--dependence that
arises through this quantity is likely to be quite robust.  On the other hand,
predicting $f_g \rho_{\mathrm{vir}}^{-1/2}$ requires a structure formation model,
and the cosmology dependence through this quantity will be necessarily model dependent.
In particular, while $\rho_{\mathrm{vir}}$ depends only on the details of nonlinear 
gravitational collapse, $f_g$ also depends on gas physics -- in particular,
in our case, on our assumption of hydrostatic equilibrium.
For each
cosmological parameter, we want to know whether these two dependencies
work in the same direction, or whether they tend to cancel each other
-- and, in either case, it is useful to know which dependence
dominates the constraints.

To answer this question, we computed $d\ln{Y}/dp$ separately from each
cosmology--dependent term, since the Fisher matrix element
(eq.~\ref{eqn:fsc_single}) is proportional to this derivative. First,
we allow cosmological parameters to vary when we compute $D_A$, but
artificially keep them at their fiducial values in the computation of
$f_g$ and $\rho_{\mathrm{vir}}^{-1/2}$. The resulting $d\ln{Y}/dp$
quantifies the dependence through $D_A^{-2}$ alone. We next allow
cosmological parameters vary in $f_g$ and
$\rho_{\mathrm{vir}}^{-1/2}$, but we keep them fixed in $D_A$; this
yields the dependence through $f_g \rho_{\mathrm{vir}}^{-1/2}$
alone. \footnote{Although we will keep referring to the
combination $f_g \rho_{\mathrm{vir}}^{-1/2}$, it is useful to clarify
that for $w$, the dependence through $\rho_{\mathrm{vir}}$ is always
much stronger than the mild cosmology--dependence through $f_g$
arising from eq.~(\ref{eqn:equilibrium}).  On the other hand, we find
that $f_g$ is much more sensitive to $\Omega_m$ and $\Omega_b$ than
$\rho_{\mathrm{vir}}$.}  The overall dependence $d\ln{Y}/dp$ is simply
the sum of these two. We compute the above derivatives at $z=0.2$ and
$z=1.5$ and the mass is set to $10^{15}\msun$ (we did not find a
strong mass dependence in the derivatives, so these numbers are
typical for clusters in the whole mass range of interest).

The results of the above exercise are listed in Tables
\ref{tbl:dependence1} and \ref{tbl:dependence2}. As we can see from
these two Tables, for each of the cosmological parameters, the
derivative through $D_A^{-2}$ and $f_g \rho_{\mathrm{vir}}^{-1/2}$ have
different signs (except for $\Omega_b$, to which $D_A$ has no
sensitivity). This, unfortunately, means that the dependence from
$D_A^{-2}$ always cancels with the dependence from $f_g
\rho_{\mathrm{vir}}^{-1/2}$. For $\Omega_m$ and $\Omega_b$, the overall
derivative is driven overwhelmingly by $f_g \rho_{\mathrm{vir}}^{-1/2}$,
and correspondingly the constraints come through $f_g
\rho_{\mathrm{vir}}^{-1/2}$. For $\Omega_{DE}$ and $h$, there are
significant cancellations, and the overall derivative has the same sign
as that from $D_A^{2}$, so the constraints come predominantly through
$D_A^{-2}$. For $w_0$ and $w_a$, there are again significant
cancellations, and the overall constraints come predominantly through
$f_g \rho_{\mathrm{vir}}^{-1/2}$ at low redshift, but through $D_A^{-2}$
at high redshift.

Although each parameter has a dependence through $f_g
\rho_{\mathrm{vir}}^{-1/2}$, the situation is different for ($\Omega_m$,
$\Omega_b$) and for ($w_0$, $w_a$). $\Omega_m$ and $\Omega_b$ are both
directly related to $f_g$, while they have a smaller or no effect on
$\rho_{\mathrm{vir}}$. The gas fraction $f_g$ is roughly proportional to
the global baryon fraction $f_b=\Omega_b/\Omega_m$. This direct
dependence is the strongest among all dependencies of $Y-T$ on
cosmological parameters. The dark energy equation--of--state
parameters $w_0$ and $w_a$, on the other hand, have no direct effect
on $f_g$, and they mainly come into play through $\rho_{\mathrm{vir}}$.
A higher $w$ induces a higher $\rho_{\mathrm{vir}}$, because clusters
collapse earlier in such a universe (Kuhlen et al. 2005). Higher
density means higher temperature for given cluster mass
(eqs.~\ref{eqn:explain2_1} and \ref{eqn:explain2_2}), or conversely,
lower mass for a given temperature.  As a result, $Y$ is reduced for a
given $T$ (eq. \ref{eqn:explain1}). This reduction is further enhanced
by the indirect effect of $\rho_{\mathrm{vir}}$ on $f_g$ through the
pressure boundary condition. This can be understood by recalling that
the ICM is assumed to be confined by the external pressure of the
infalling gas. This pressure is proportional to $\Omega_b T/\Omega_m$
(eqs.~\ref{eqn:boundary} and \ref{eqn:explain2_1}). When the virial
density is raised, the external pressure increases roughly linearly
with the temperature, which implies that the gas density is
approximately kept constant ($P\propto \rho T$). The gas fraction,
which is roughly proportional to the ratio of gas density to the
virial overdensity, is therefore reduced.

\begin{table}
  \caption{$d\ln{Y}/dp$ evaluated at a fixed cluster temperature at $z=0.2$, where $p$ is any of the 6 cosmological parameters.  The three columns, from left to right, show the values when we include the dependence only via $D_A^{-2}$, or only via $f_g \rho_{\mathrm{vir}}^{-1/2}$, or the full derivative.}
    \label{tbl:dependence1}
\begin{center}
\begin{tabular}{ c c c c}
\hline \hline
Parameter & $D_A^{-2}$ &  $f_g \rho_{\mathrm{vir}}^{-1/2}$ & Overall\\
\hline
$\Omega_m$ & +0.13 & -6.20 & -6.07\\
$\Omega_{DE}$ & -0.19 & +0.15 & -0.03\\
$\Omega_b$ & 0 & +27.50 & +27.50\\
$ h$ & +2.59 & -1.34 & +1.25\\
$ w_0$ & +0.19 & -0.38 & -0.19 \\
$ w_a$ & +0.01 & -0.05 & -0.04 \\
 \hline
\end{tabular}
\end{center}
\end{table}

\begin{table}
  \caption{Same as Table \ref{tbl:dependence1} except for $z=1.5$.}
    \label{tbl:dependence2}
\begin{center}
\begin{tabular}{ c c c c}
\hline \hline
Cosmological parameters & $D_A^{-2}$ &  $f_g \rho_{\mathrm{vir}}^{-1/2}$ & Overall\\
\hline
$\Omega_m$ & +0.11 & -6.75 & -5.64\\
$\Omega_{DE}$ & -0.76 & +0.68 & -0.08\\
$\Omega_b$ & 0 & +26.96 & +26.96\\
$ h$ & +2.59 & -1.37 & +1.23\\
$ w_0$ & +0.45 & -0.15 & +0.30 \\
$ w_a$ & +0.11 & -0.07 & +0.03 \\
 \hline
\end{tabular}
\end{center}
\end{table}

Among the cosmological parameters, the $Y-T$ relation is most
sensitive to $\Omega_m$ and $\Omega_b$, and constraints on these two
parameters are therefore the tightest. However they also suffer the
most from the severe degeneracy between one another (see column
$\cal{D}$ in Table \ref{tbl:constraints1}). This is because both
parameters affect the $Y-T$ relation in an approximately uniform way,
insensitive to redshift and cluster mass. This also means that a prior
on one of these two parameters from another measurement could greatly
help constrain the other.  For example, if we apply a prior of
$0.0015$ from the result of {\sl WMAP}+BAO+SN on $\Omega_b$ (Dunkley
et al. 2008), $\Delta \Omega_m$ is reduced by more than a factor of 3,
while the constraints on other parameters change only mildly. From the
simple argument that both $\Omega_m$ and $\Omega_b$ are constrained
through $f_g$ and $f_g$ is roughly proportional to the baryon fraction
$\Omega_b/\Omega_m$, one could conclude that it is the combination
$\Omega_b\Omega_m^{-1}$ that is being constrained. This conclusion is
borne out by our numerical results, which show that the direction of
the Fisher eigenvector in the $\Omega_b - \Omega_m$ subspace is in the
direction $\Omega_b\Omega_m^{-1.2}=const$; i.e., the scaling relation
indeed best constrains this combination.  Other parameters have a
comparatively smaller effect on the $Y-T$ relation (see the
``Overall'' column in Tables~\ref{tbl:dependence1} and
\ref{tbl:dependence2} and the corresponding column $(F_{ii})^{-1/2}$
in Table \ref{tbl:constraints1}). But due to the sensitivity to each
parameter having a different redshift dependence, they suffer much
weaker degeneracies. The parameter with the lowest degeneracy
(smallest $\cal{D}$ parameter) is $w_0$, with $\cal{D}$$=3.6$, less than
1/50th of the degeneracy between $\Omega_m$ and $\Omega_b$.

\begin{table*}
  \caption{Constraints from number counts approach with different scatter and different number of $Y$ bins. Columns labeled with ``Fiducial'' contain results with 10\% scatter and 8 $Y$ bins; columns labeled with ``1 $Y$ bin'' contain results with 10\% scatter and 1 $Y$ bin; columns labeled with ``No Scatter'' contain results with no scatter and 8 $Y$ bins.}
    \label{tbl:comparison}
\begin{center}
\begin{tabular}{ c c c c c c c c c c}
\hline \hline
Parameter constraints & \multicolumn{3}{c}{Fiducial} & \multicolumn{3}{c}{1 $Y$ bin}& \multicolumn{3}{c}{No Scatter}\\
 & $\sigma$& $(F_{ii})^{-1/2}$ & $\cal{D}$& $\sigma$& $(F_{ii})^{-1/2}$ & $\cal{D}$&$\sigma$&$(F_{ii})^{-1/2}$ & $\cal{D}$\\
\hline
$ \Omega_m$ & 0.023 & 0.0024 & 9.6&0.10&0.0026&39.4&0.018&0.0022&8.1\\
$ \Omega_{DE}$ &  0.29 & 0.016 & 18.4 &1.0&0.016&63.0&0.19&0.015&12.5\\
$ \Omega_b$ &  0.007 & 0.00035 & 20.2 &0.042&0.00043&99.2&0.0019&0.00030&6.4\\
$ h$ & 0.11 & 0.006 & 18.6 &0.54&0.007&75.9&0.028&0.0049&5.6\\
$ w_0$ & 0.20 & 0.018 & 11.1 &0.42&0.019&22.5&0.12&0.017&7.0\\
$ w_a$ & 1.4 & 0.10 & 14.4 & 8.1&0.10&80.3&1.0&0.10&9.9\\
$ \sigma_8$ &  0.016 & 0.0011 & 14.2 &0.07&0.0011&66.5&0.011&0.0011&10.3\\
$ n_s$ & 0.13 & 0.023 & 5.7  & 1.1&0.037&28.6&0.06&0.022&2.9\\
 \hline
\end{tabular}
\end{center}
\end{table*}

Analogous to the discussion above, previous works have clarified the
cosmology--dependence of the cluster number counts. We refer the
reader to, e.g., Haiman et al. (2001) for a detailed discussion; here
we just emphasize two points. First, the number counts also include a
cosmology--dependence from the $Y-M$ relation, through the selection
function $g(Y_{\alpha}, M)$. The number--count constraint on $\Omega_b$
is driven through this dependence, but the constraints on other
parameters are dominated by either the cosmological volume element or
the growth function (except the $\Omega_m$ dependence, which is
dominated by the explicit linear scaling of the cluster mass function
with $\Omega_m$).  Second, we not only have multiple redshift bins,
but also multiple $Y$--bins.  This helps significantly in constraining
cosmological parameters (analogously to the shape of the cluster mass
function being helpful; e.g., Hu 2003).  In Table~\ref{tbl:comparison}
below, we present constraints both with and without binning in
$Y$. Comparing these two cases, we see that the $\cal{D}$ parameter
changes significantly, while $(F_{ii})^{-1/2}$ has only a mild
change. This shows that the tightening of the constraints in the case
when 8 $Y$ bins are used is achieved mainly via breaking degeneracies
between different parameters. In the same table, we also list
constraints when the scatter between $Y$ and $M$ is set to zero. The
constraints become more stringent, which highlights the degrading
effect of the $Y-M$ scatter through flattening the mass function. A
small scatter in the mass--observable relation (10\% as assumed in
this work) is seen to degrade constraints by a factor of up to 4 (the
parameter $h$). Again, we see that this is mainly due to higher
degeneracies in the presence of the scatter.

%\newpage
%%%%%%%%%%%%%%%%%%%%%%%%%%%%%%%%%%%%%%%%%%%%%%%%%%%%%%%%%%%%%%%%%%%%%%%%%%%%%%%%
\subsection{Constraints with cosmological and cluster parameters}
\label{subsec:results_2}
%%%%%%%%%%%%%%%%%%%%%%%%%%%%%%%%%%%%%%%%%%%%%%%%%%%%%%%%%%%%%%%%%%%%%%%%%%%%%%%%

Below, we consider the more ``realistic'' case, in which we take into
account uncertainties in cluster structure and its evolution. In
\S~\ref{subsec:model}, we have parameterized cluster structure and
evolution with 15 parameters, characterizing various aspects of ICM
physics, namely the shape of the gravitational potential, the gas
entropy, non-thermal pressure, and boundary condition, as well as the
mass-- and redshift--dependence of these parametrized quantities.  We
repeat the above analysis, but also including these cluster structure
parameters, which means we constrain 23 parameters simultaneously. The
results of this exercise are shown in Table
\ref{tbl:constraints2}. Within the parentheses next to the errors on
the cosmological parameters, we list the ratio of errors,
$\cal{R}$=$\sigma_2/\sigma_1$, where $\sigma_1$ is the idealized
constraint shown in Table~\ref{tbl:constraints1}, and $\sigma_2$ is
the new value in Table~\ref{tbl:constraints2}. This ratio $\cal{R}$
therefore quantifies the degradation of the constraint introduced by
the cluster parameter uncertainties. For the majority of parameters,
the degradation is less than a factor of 2, and the constraints remain
tight, despite the large increase in parameter space (this is true for
both the scaling relation and number counts approaches).  We emphasize
again that we simplified things by assuming a particular form
(eqs.~\ref{eqn:parameter}) for the mass and redshift dependence of
cluster parameters; nevertheless, these results highlight the ability
of upcoming surveys to constrain a large number of parameters, which
is due essentially to the large number of clusters and therefore small
statistical errors.

For the number counts approach, the largest degradation is on
$\Omega_b$. This is understandable, because unlike for other
cosmological parameters, the constraint on $\Omega_b$, as we have
mentioned, is largely from the $Y-M$ relation in eq.~(\ref{eqn:g}),
not from the mass function itself, and all of the 15 cluster
parameters affect the $Y-M$ relation. This could account for the large
degeneracy between $\Omega_b$ and the cluster parameters.  Of course,
$\Omega_b$ is measured accurately by other methods (Kirkman et
al. 2003; Dunkley et al. 2008), and this degradation is not a concern.
In the scaling relations, however, the largest degradation is suffered
by $w_0$. This is because the simple power--law parameterization in
eq.~\ref{eqn:parameter} happens to be close to the way $w_0$ affects
the evolution of the scaling relation.  This large degeneracy between
the DE equation--of--state parameters and cluster parameter suggests
that the cluster constraints on $w_0$ and $w_a$ will be especially
useful when combined with independent measurements of these parameters
using other probes.

Table~\ref{tbl:constraints2} shows further that the constraints on
cluster parameters are, in general, quite weak from both approaches,
with most constraints at the order--unity level. This, conversely,
indicates that the $Y$ parameter is relatively insensitive to cluster
parameter variations.  We see from Table~\ref{tbl:constraints2} that
the single--parameter errors $(F_{ii})^{-1/2}$ for the cluster
parameters are very low, showing that the weakness of these
constraints are due to strong degeneracies among the cluster model
parameters.  One likely reason for this strong degeneracy is that we
adopted the same power--law form for the mass-- and
redshift--evolution for each of the cluster parameters.  Indeed, we
find that introducing even unrealistically tight priors on the
cosmological parameters do not improve cluster parameter constrains
significantly, indicating that the degeneracies are among the cluster
parameters themselves.  We thus conclude that $Y-T$ relation by itself
is not a good way of placing precise constraints on individual cluster
parameters, unless the mass--dependence and redshift--evolution of the
physical parameters can be understood a--priori, and they differ
significantly from the power--law forms assumed here. Of course, the
$Y-T$ relation still delivers tight constraints on cluster--parameter
combinations, so it should be useful when combined with other cluster
observables.

\begin{table*}
\begin{center}
\caption{Constraints as in
Table~\ref{tbl:constraints1}, except in the more realistic case that
includes cluster structure parameters.  Within the parentheses next to
the errors on each cosmological parameter, we list the factor by which
the constraints degrade relative to the idealized case.}
\label{tbl:constraints2}
\begin{tabular}{c c c c c c c c}
\hline\hline
Parameter & \multicolumn{2}{c}{SR} & \multicolumn{2}{c}{NC} & \multicolumn{3}{c}{Combined}\\
&$\sigma$& $(F_{ii})^{-1/2}$&$\sigma$& $(F_{ii})^{-1/2}$&$\sigma$& $(F_{ii})^{-1/2}$&$\xi$\\
\hline
$\Omega_m$& 0.087(1.6)&0.00030 & 0.068(2.9)& 0.0024& 0.038(4.2)&0.00030&0.50\\
$\Omega_{DE}$ &0.28(1.4)&0.029 &0.34(1.2)& 0.016&0.15(2.5)&0.014&0.46\\
$\Omega_b$& 0.080(6.7)&0.000062 & 0.10(14.3)&0.00035& 0.038(10.3)&0.000061&0.38\\
$h$ & 0.12(1.5)&0.0014 & 0.14(1.3)&0.0060& 0.075(1.5)&0.0014&0.70\\
$w_0$ & 0.53(14.3)&0.010& 0.34(1.7)&0.018& 0.22(13.8)&0.0089&0.58\\
$w_a$ &0.64(3.0)&0.044& 1.63(1.2)&0.099& 0.45(4.1)&0.040&0.58\\
$\sigma_8$&N/A&N/A&0.055(3.4)&0.0011& 0.033(4.7)&0.0011&0.36\\
$n_s$& N/A& N/A&0.87(6.5)& 0.023&0.46(12.8)&0.023&0.28\\
$\hat{K}_{\mathrm{norm}}$&4.37&0.0042&2.25&0.020&0.56&0.0041&0.08\\
$\hat{K}_{\mathrm{m}}$&0.67&0.0014&0.96&0.010&0.24&0.0014&0.19\\
$\hat{K}_{\mathrm{z}}$&1.95&0.0048&2.03&0.022&0.79&0.0047&0.32\\
$s_{\mathrm{norm}}$&0.98&0.00092&2.40&0.0097&0.57&0.00092&0.40\\
$s_{\mathrm{m}}$&0.075&0.00061&1.75&0.010&0.042&0.00061&0.32\\
$s_{\mathrm{z}}$&0.96&0.0024&3.56&0.023&0.48&0.0024&0.27\\
$b_{\mathrm{norm}}$&3.54&0.0063&2.60&0.014&1.02&0.0057&0.24\\
$b_{\mathrm{m}}$&1.92&0.0050&1.83&0.015&0.73&0.0048&0.30\\
$b_{\mathrm{z}}$&0.011&0.0026&0.26&0.0093&0.0083&0.0025&0.61\\
$c^{\mathrm{NFW}}_{\mathrm{norm}}$&2.02&0.0030&6.67&0.076&1.03&0.0030&0.29\\
$c^{\mathrm{NFW}}_{\mathrm{m}}$& 1.05&0.0024&6.59&0.10&0.49&0.0024&0.22\\
$c^{\mathrm{NFW}}_{\mathrm{z}}$&2.47&0.0085&5.74&0.25&1.33&0.0085&0.35\\
$\eta_{\mathrm{norm}}$&0.85&0.0016&0.45&0.0050&0.18&0.0015&0.20\\
$\eta_{\mathrm{m}}$&0.97&0.0015&0.85&0.0062&0.24&0.0015&0.14\\
$\eta_{\mathrm{z}}$&1.09&0.0049&1.33&0.013&0.47&0.0046&0.31\\
 \hline 
\end{tabular}
\end{center}
\end{table*}

%%%%%%%%%%%%%%%%%%%%%%%%%%%%%%%%%%%%%%%%%%%%%%%%%%%%%%%%%%%%%%%%%%%%%%%%%%%%%%%%
\subsection{Effects of scatter and completeness uncertainties}
\label{subsec:results_3}
%%%%%%%%%%%%%%%%%%%%%%%%%%%%%%%%%%%%%%%%%%%%%%%%%%%%%%%%%%%%%%%%%%%%%%%%%%%%%%%%
 
In addition to the uncertainties in cluster structure, there are also
uncertainties in scatter and completeness. The scaling relation test
is affected by scatter and incompleteness only indirectly through
Malmquist bias.  This bias is the increase in the mean value of $Y$ at
fixed $T$, because the lowest--$Y$ clusters that are scattered below
the detection threshold are missing from the sample.  We find that the
scatter changes the number of detectable clusters in each redshift bin
by less than 1\% (except in the three bins beyond $z=1.8$, where the
changes are between 1-2\%), and the mean $\langle Y\rangle$ is changed
by a similar amount. This is comparable to the change in $\langle
Y\rangle$ caused by variations in our parameters within their
marginalized $1\sigma$ uncertainties (see
Tables~\ref{tbl:dependence1}, \ref{tbl:dependence2} and
\ref{tbl:constraints1}).  However, since flux limited surveys can do a
correction that should eliminate the bulk of the Malmquist bias, we
believe this will not be a major limitation of the constraints.

The effects of scatter and completeness uncertainties on the number
counts constraints is somewhat more subtle. In this section, we allow
both the scatter and a completeness to vary, together with the
cosmological and cluster parameters. The scatter between $M$ and $Y$
is parameterized using the same power--law form as the cluster
parameters (eq. ~\ref{eqn:parameter}). The completeness $\cal{C}$,
defined as the fraction of clusters at a given $Y$ at redshift $z$
that are detected is taken to be given by a similar power--law, except
$M/M^{\star}$ is replaced by $Y/Y^{\star}$ in
equation~(\ref{eqn:parameter}), where $Y^{\star}$ is chosen to be
$1$mJy.

The fiducial scatter is assumed to be 10\%, and the fiducial
completeness is set to be 100\% (both independent of redshift and
mass, except completeness is set to zero below $M_{\rm min}$).  Table
\ref{tbl:constraints3} shows the results when scatter is included and
when both scatter and incompleteness are included. For comparison, we
also list the result when neither uncertainty is taken into account
(repeating the NC column from Table~\ref{tbl:constraints2}).  The
degradations are relatively small for most parameters. The two
exception are $\Omega_m$ and $\sigma_8$, for which the number-count
constraints degrade by a factor of $\sim 3$ when the completeness
uncertainty is included.  The impact of the completeness uncertainty
is relatively modest, because our treatment assumes that we know the
form of the dependence of $\cal{C}$ on $Y$ and $z$ (i.e. power--laws
in our case).  At the opposite extreme, if one allows completeness to
be an arbitrary function of mass and redshift, then of course no
constraint can be derived on any model parameter.  The fact that we
still find interesting constraints shows that a reliable
parameterization of the completeness as a function of $Y$ and $z$ will
be very important.  We also find that all of the constraints would
recover their values (to within 30\%) in the fixed $\cal{C}=$1 case
when a prior of 15\% is applied to $\cal{C}$.  Finally, in
Table~\ref{tbl:constraints3}, we also list the combined constraints
from the scaling relations and the number counts.  Comparing these
values with the combined constraints listed in Table
\ref{tbl:constraints2}, we find that these constraints are less
affected by incompleteness than those from number counts approach
alone. Except for $\sigma_8$, which degrades by a factor of $\sim 2$,
the constraints all degrade by factors of $\lsim 1.5$.

\begin{table}
\caption{Constraints from number counts approach with scatter and incompleteness taken into account. First column is the result when the uncertainties on scatter and completeness are ignored; second column is the result when the uncertainty on scatter is included; third column is the result when both uncertainties are taken into account; lastly we also list the combined constraints from number counts approach and scaling relation approach with scatter and incompleteness included.}
\label{tbl:constraints3}
\begin{tabular}{c c c c c c}
\hline\hline
Parameter & None & +S$^{\dagger}$& +S+I$^{\dagger}$& \multicolumn{2}{c}{+S+I$^{\dagger}$} \\
    & NC & NC & NC & Combined &$\xi$\\
\hline
$\Omega_m$&0.068 &0.072 &0.225 & 0.060&0.55\\
$\Omega_{DE}$ &0.34 &0.35 &0.66 & 0.22&0.72\\
$\Omega_b$& 0.10&0.12 &0.12 & 0.053&0.65\\
$h$ & 0.14&0.17 & 0.18 & 0.081&0.68\\
$w_0$ &0.35&0.46& 0.56 & 0.27&0.49\\
$w_a$ & 1.63&1.66 &1.78 & 0.48&0.66\\
$\sigma_8$&0.055& 0.061&0.17 & 0.072&0.19\\
$n_s$&0.87& 0.92&1.08& 0.64&0.35\\
$\hat{K}_{\mathrm{norm}}$&2.25&2.73&2.78&1.41&0.36\\
$\hat{K}_{\mathrm{m}}$&0.96&1.24&1.48&0.26&0.18\\
$\hat{K}_{\mathrm{z}}$&2.03&2.08&2.12&0.91&0.40\\
$s_{\mathrm{norm}}$&2.40&2.90&2.93&0.68&0.54\\
$s_{\mathrm{m}}$&1.75&1.88&2.07&0.045&0.36\\
$s_{\mathrm{z}}$&3.56&3.77&3.99&0.52&0.31\\
$b_{\mathrm{norm}}$&2.60&3.11&3.16&1.16&0.24\\
$b_{\mathrm{m}}$&1.83&2.17&2.33&0.86&0.34\\
$b_{\mathrm{z}}$&0.26&0.44&0.45&0.0089&0.70\\
$c^{\mathrm{NFW}}_{\mathrm{norm}}$&6.67&7.25&7.66&1.45&0.55\\
$c^{\mathrm{NFW}}_{\mathrm{m}}$&6.59&6.99&7.42&0.57&0.30\\
$c^{\mathrm{NFW}}_{\mathrm{z}}$&5.74&6.08&7.18&1.38&0.35\\
$\eta_{\mathrm{norm}}$&0.45&0.64&0.64&0.26&0.26\\
$\eta_{\mathrm{m}}$&0.85&0.93&0.96&0.26&0.14\\
$\eta_{\mathrm{z}}$&1.39&1.38&1.44&0.48&0.30\\
$\sigma_{\mathrm{norm}}$&&0.18&0.18&0.088&0.23\\
$\sigma_{\mathrm{m}}$&&7.60&7.79&3.77&0.23\\
$\sigma_{\mathrm{z}}$&&8.64&8.77&5.75&0.43\\
$\cal{C}_{\mathrm{norm}}$&&&1.72&0.56&0.11\\
$\cal{C}_{\mathrm{m}}$&&&0.13&0.088&0.46\\
$\cal{C}_{\mathrm{z}}$&&&2.68&1.14&0.18\\
 \hline
\end{tabular}

\raggedright $^{\dagger}$ ``S'' stands for ``Scatter'', ``I'' stands for ``Incompleteness''.
%\end{center}
\end{table}

The physical origin of scatter in the observables $Y$ and $T$ can
be cluster-to-cluster variations in the structure parameters (even in
the context of our idealized spherical models), and geometrical
effects (i.e. viewing aspherical clusters along different
sight--lines).  As mentioned in \S~\ref{subsec:model} above, in
addition to the Malmquist bias, underlying scatter in a physical
structure parameter can cause a bias by producing a skewed probability
distribution for $Y$.  In order to assess how large this additional
bias might be, we have performed the following calculation.  First,
looking at the third column in Table~\ref{tbl:constraints2}, we see
that the best constrained individual cluster parameter is
$s_{\mathrm{norm}}$. This suggests that among the cluster structure
parameters, it is this parameter (the slope of the entropy profile)
that could cause the largest bias.  We then assumed a symmetric,
Gaussian scatter on this parameter, with an r.m.s. equal to 10\% of
its fiducial value ($\sigma_{s_{\rm norm}} = 0.11$).  We then computed
the distribution of $Y$-values, for clusters at $z=0.5$, and $T=7$keV,
induced by this Gaussian scatter\footnote{Note that changes in
$s_{\mathrm{norm}}$ change both $Y$ and $T$; since we are interested
in the distribution of $Y$ at fixed $T$, we adjust the cluster mass
$M$, for each value of $s_{\rm norm}$, to keep $T$ fixed at 7 keV.}.
In the absence of any scatter in $s_{\rm norm}$, the mean SZ decrement
at $T=7$keV is $\langle Y\rangle=70.46$mJy.  When the scatter is
included, we find that $\langle Y\rangle$ is increased by 2\%, to
71.88 mJy.

This level of bias is certainly a concern. From the parameter
constraints $\sigma(p)$ listed in Table~\ref{tbl:constraints1}, and
the logarithmic dependencies of $Y$ vs $p$ listed in
Tables~\ref{tbl:dependence1} and \ref{tbl:dependence2}, we infer that
systematic errors on the measurement of $\langle Y\rangle $ at fixed
$T$ should be controlled to within $\approx 1\%$, in order for the
corresponding bias on the parameters not to exceed the $1\sigma$
constraints.  Furthermore, in the above exercise, we find a standard
deviation of $\sigma(Y)=14~{\rm mJy}=0.194 \langle Y\rangle$; i.e. a
value that is almost double the fiducial adopted 10\% scatter in the
$Y-T$ relation.  These numbers suggests that, in order to realize
constraints comparable to the forecasts we present here, it will be
necessary to obtain a physical understanding of the scatter in
cluster-structure from hydrodynamical simulations.  When the technique
proposed here is applied to an actual large data--set, it will be
important to include parameters that describe such scatter. The full
distribution of $Y$ at fixed $T$, rather than the single value
$\langle Y \rangle$, can then be used as additional signal, in order
to constrain these extra parameters.

%%%%%%%%%%%%%%%%%%%%%%%%%%%%%%%%%%%%%%%%%%%%%%%%%%%%%%%%%%%%%%%%%%%%%%%%%%%%%%%%
\section{Discussion}
\label{sec:discussion}
%%%%%%%%%%%%%%%%%%%%%%%%%%%%%%%%%%%%%%%%%%%%%%%%%%%%%%%%%%%%%%%%%%%%%%%%%%%%%%%%

\subsection{Comparing the $Y-T$ and $dN/dz$ Constraints}
\label{subsec:discussion_compare}

We have forecast the constraints from $Y-T$ scaling relation approach
and number counts approach. Interestingly, we find that these two
approaches yield comparable results, even though the cluster abundance
is more sensitive to cosmological parameters than the scaling
relation.  There are several reasons, however, for the scaling
relations to be competitive, at least statistically.  First, the
number counts only utilize $Y$ and $z$ of a cluster, while the scaling
relation also derives information from the temperature $T$.  Second,
in the scaling relations, each cluster adds a new data point, so that
we are effectively using $N\approx 6,800$ independent measurements of
the observable $Y$. Each $Y$--measurement has a fractional error
significantly below order unity (of order $\Delta Y/Y=\sigma_{Y,T}
\approx 20$\% in eq.~\ref{eqn:fsc_single}), so that when these are
combined independently, the effective combined statistical error is
$\sigma_{Y,T}/\sqrt{N}$.  This is better than the total effective
Poisson error ($\sim \sqrt{N}$) from the cluster counts.  This
  statement holds as long as $Y$ is measured at fixed $T$ with an
  uncertainty better than order unity, since effectively, in the
  number count approach, each cluster contributes an order--unity
  statistical error.  This comparison neglects systematic errors,
  which can ultimately limit the constraints from both approaches.  In
  particular, systematic errors in both the measurements of $Y(T)$ and
  its model predictions have to be at the $\lsim 1/\sqrt{N} \approx
  1\%$ level, in order not to dominate over the statistical errors;
  likewise, selection effects for the cluster counts have to be
  controlled to this level of systematic accuracy.
Third, unlike cluster counts, the scaling relation does not
explicitly depend on $\sigma_8$ and $n_s$. The only dependence is
through the Malmquist bias, but we find this dependence to be
negligibly small, even when varying $\sigma_8$ in the range of
0.7-0.9.  Therefore, the scaling relation avoids degeneracies
involving these parameters. Finally, as mentioned above, the number
counts approach is less robust to selection errors.
 
Quantitatively, the scaling relation approach yields tighter
constraints on $\Omega_{DE}$, $h$, $w_0$ and $w_a$ when the
uncertainty in cluster structure is ignored, while the number counts
do better for the other parameters. Once we include cluster structure
uncertainty, the constraint on $w_0$ from the scaling relations and
$\Omega_b$ from the number counts suffer bad degradations. 

\subsection{Combined $Y-T$ and $dN/dz$ Constraints}
\label{subsec:discussion_compare}

We also computed the combined constraints from the two methods.  
  As stated above, we assume that the two constraints are independent,
  and simply add the Fisher matrices; we justify this assumption in
  \S~\ref{subsec:discussion_covariance} below. Combining the scaling relation with
the number counts is useful to break degeneracies. For example, in the
idealized case, the combined constraint on $n_s$ is a factor of 3
tighter than that from number counts alone. This improvement is
entirely from degeneracy breaking, since the scaling relation approach
does not directly constrain $n_s$.  While the $n_s$ constraints in the
realistic case degrade to uninteresting levels, the degeneracy
breaking is also helpful in constraining other parameters. To quantify
this, we computed the ``complementarity'' parameter, which defined as
(Fang \& Haiman 2008),
\begin{eqnarray}
\xi=(\Delta p^{\mathrm{combined}})^{2}\left[\frac{1}{(\Delta p^{\mathrm{sr}})^2}+\frac{1}{(\Delta p^{\mathrm{nc}})^2}\right]
\label{eqn:comparameter}
\end{eqnarray}
In the idealized case (Table~\ref{tbl:constraints1}), this parameter
is below 0.2 for 5 of the 8 cosmological parameters, meaning that for
these parameters, the combined constraints are more than a factor of
two tighter than simply adding the two results in quadrature. This
parameter is larger in the realistic case
(Table~\ref{tbl:constraints2}), with an average of $\sim 0.48$ for the
cosmological parameters (the lowest value is 0.28, for $n_s$).  

\subsection{Beyond Power-Law Cluster Structure Models}
\label{subsec:discussion_curvedmodels}

In Figure~\ref{fig:lt}, we saw a hint that our power--law
parameterizations (in eq.~\ref{eqn:parameter} might be insufficient
over an extended mass (and possibly also redshift) range. Since the
degeneracy between cosmological and cluster parameters depends on this
explicit form, we investigated the impact of allowing ``curvature'' in
these power--laws. Specifically, we modified
equation~(\ref{eqn:parameter}) to include higher order terms,
\begin{eqnarray}
\label{eqn:parameter2}
\ln{p}&=&\ln{p_{\mathrm{norm}}}+p_{m1}\ln{\frac{M}{M^{\star}}}+p_{m2}\left(\ln{\frac{M}{M^{\star}}}\right)^2\\
\nonumber
      & &+p_{z1}\ln{(1+z)}+p_{z2}\left[\ln{(1+z)}\right]^2.
\end{eqnarray}
With the inclusion of these new terms, we have $5\times 5=25$ cluster
model parameters. We use the Fisher matrix technique to forecast
constraints on these 25 parameters, together with the cosmological
parameters. Compared to the 15 cluster parameter case (Table
\ref{tbl:constraints2}), we find that the constraints on $\Omega_m$,
$w_0$ and $w_a$ are the most affected, with an increase in their
marginalized errors by a factor of $\approx 2$. From 
%column $\cal{R}$
the large degradation factors ($\sim 10$) shown in the parentheses in
Table~\ref{tbl:constraints2}, we see that $\Omega_b$ is sensitive to
cluster structure uncertainties, as well. However, we find that the
constraints on $\Omega_{DE}$ and $h$ are relatively robust to these
uncertainties. The reason is that, as shown above, the constraints on
these two parameters arise through $D_A$, which is not affected by
cluster parameters, while the constraints on all other parameters
receive a significant contribution from the cluster properties. We
also find that if we simultaneously impose a prior of 50\% on the
fractional error on each of the 25 cluster parameters, this recovers
constraints similar to that of 15 cluster parameters case.  This
suggests that relatively weak priors may mitigate the impact of more
complicated cluster structure models.

\subsection{Uncertainties from the Low--Redshift Mass--Floor}
\label{subsec:discussion_massfloor}

At low redshifts, the mass corresponding to a simple constant SZ flux
is very low -- dropping below masses corresponding to galaxy
clusters. The nearby objects extend a large solid angle, so that
surface brightness selection effects are no longer negligible. While
this issue can be addressed by using different cluster--finding
algorithms (e.g., Sehgal et al. 2007), for simplicity, we have imposed
a constant mass limit of $10^{14}h^{-1} \msun$.\footnote{The
  selection will also be affected by instrumental specifications, such
  as the beam profile, which will have to be taken into
  account in an actual analysis.}  To see whether the constraints are
sensitive to the (somewhat arbitrary) choice of this mass floor, we
re--computed our constraints for a different value of
$5\times10^{13}h^{-1} \msun$.  Lowering the mass floor increases the
total number of clusters by $\sim 13$\% (i.e. by $\sim 900$ new clusters
at redshift below $\sim 0.2$, where the mass floor is above
  the mass limit set by the flux threshold). The scaling relation
results are not very sensitive to this change; the difference in the
constraints is only a few percent in the idealized case, and up to
20\% in the realistic case. The results, however, change more
significantly in the number counts approach. In the idealized case,
$\Omega_m$, $\Omega_{DE}$, $w_0$ and $w_a$ are most affected -- their
constraints improve to $0.014$, $0.080$, $0.094$ and $0.28$
respectively (compared to the original results of $0.023$, $0.29$,
$0.20$ and $1.4$). These large improvements cannot be explained by the
modest, $\sim13$\% increase in the number of clusters, implying that
the low--redshift, low--mass clusters help break the significant
degeneracies among these parameters. In the realistic case, lowering
the mass floor by a factor of two still improves the constraints on
($\Omega_m$, $\Omega_{DE}$, $w_0$, $w_a$) by factors of (1.4, 1.9,
1.6, 4.6).  This underscores the importance of an accurate measurement
of the abundance of low-mass clusters at low redshifts.  It also shows
that mis--estimates in the value of the mass floor can potentially
introduce a large bias; such mis--estimates are equivalent to errors
in the selection function as discussed in \S~\ref{subsec:results_3}
above.

\subsection{Uncertainties from Excising the Cluster Core}
\label{subsec:discussion_core}

A potential concern is that in our $Y-T$ scaling relation, we
restricted the definition of the emission--weighted temperature
outside the radius $0.15R_{500}$ (see eq.~\ref{eqn:tew}).  While this
eliminates the sensitivity of the temperature to the complex physics
in the cluster core, it can still introduce uncertainties or a bias in
the inferred $T_{\mathrm{ew}}$, since the inner radius has to be
estimated from the data itself.  We can ask the simple question: how
accurately do we have to know $R_{500}$, in order for the
corresponding bias not to exceed the $1\sigma$ constraints we obtained
above?  As mentioned in \S~\ref{subsec:results_3} above, the
systematic errors on the measurement of $Y$ at fixed $T_{\rm ew}$
should be controlled to within $\approx 1\%$. Since $Y\propto T_{\rm
ew}^{2.5}$ (see eq. \ref{eqn:explain3}), this implies that the
systematics of the $T_{\rm ew}$ measurement has to be accurate to
$\approx 0.4\%$.  We emphasize that this is the required systematic
error, and a much larger scatter on the measurements for individual
clusters is tolerable (in our case, the individual temperatures need
to be known to an accuracy of $\sqrt{6,800}\times 0.4\%\approx 30\%$).
We find that a 0.4\% change in the temperature corresponds to a 5\%
change in the radius (i.e. in the lower limit of the integral in
eq.~\ref{eqn:tew}; for reference, the difference between $R_{600}$ and
$R_{500}$ is around 8\%). The cosmological dependence of $R_{500}$ is
smaller than this: a $1\sigma$ (=0.055) change in $\Omega_m$ induces a
3.5\% change in $R_{500}$, and other cosmological dependencies are
much weaker. We conclude that the cosmology--dependence of $R_{500}$
will not degrade the errors by more than $1\sigma$.  However, the 5\%
error in the radius, translates into a 15\% systematic error
requirement on the enclosed mass.  It should be possible to calibrate
the $Y-M$ relation to within 15\% systematic accuracy (Kravtsov et
al. 2006), especially if $\sim 10\%$ priors on $\Omega_m$ and
$\Omega_b$ are used.

\subsection{Using Priors on Cluster Structure to Improve Constraints}
\label{subsec:discussion_priors}

In the realistic case above, we have arguably been pessimistic by
allowing all of the cluster parameters to vary arbitrarily. In
reality, useful priors may be available on these parameters, either
from other observations, or from simulations. For example, detailed
X--ray measurements already reveal ICM entropy profiles in low
redshift clusters (Ponman et al. 2003, Pratt. et al. 2006). Likewise,
the NFW concentration parameter has been carefully studied in
numerical simulations (e.g., Navarro et al. 1997; Wechsler et al. 2002;
Kuhlen et al. 2005). To assess whether such priors could improve
constraints on cosmological parameters, in Table
\ref{tbl:constraints6} we present the results of calculations that
adopt a prior of 1.0 and 0.1 on the fractional errors on all cluster
parameters.  Arguably, order--unity priors should be achievable, and
in this sense, the results in Table \ref{tbl:constraints6} are even
more ``realistic'' than those without priors.  While improvements are
noticeable (approaching factors of two for $w_0$ and $w_a$) for the
SR--only constraints even with these weak priors,
Table~\ref{tbl:constraints6} shows that improving the combined SR+NC
constraints by a factor of two requires $\sim 10$\% priors.

In order to assess whether a single cluster parameter is the
``culprit'', we next tried applying priors on individual cluster
parameters.  We found no significant improvements, even if we applied
priors of 0.001 on any one of the cluster parameters. This shows that
the degeneracies are all inherently multi--dimensional.  As an another
series of exercises, we applied simultaneous priors either on the set
of all normalization parameters ($\hat{K}_{\mathrm{norm}}$,
$s_{\mathrm{norm}}$, $b_{\mathrm{norm}}$,
$c^{\mathrm{NFW}}_{\mathrm{norm}}$, $\eta_{\mathrm{norm}}$), the set
of all mass--dependence parameters, or the set of all
redshift--dependence parameters.  We focus on how these priors affect
SR constraints on $w_0$, since this is the constraint most affected by
the cluster parameters. We find that the largest improvements are
afforded by priors on redshift--dependence parameters. The constraint
improves from 0.53 to 0.25 when a prior of 0.1 is applied to all
redshift dependence parameters.

\begin{table*}
  \caption{Constraints from the scaling relations, using pessimistic or optimistic priors on cluster model parameters. Priors are applied simultaneously to all of the cluster parameters, but only the constraints on cosmological parameters are listed.}
    \label{tbl:constraints6}
\begin{center}
\begin{tabular}{ c c c c c c c c c}
\hline \hline
Para-- & \multicolumn{4}{c}{Prior of 1} & \multicolumn{4}{c}{Prior of 0.1}\\
meter & SR & NC & SR+NC&$\xi$ & SR & NC & SR+NC&$\xi$\\
\hline
$ \Omega_m$ & 0.071& 0.053 & 0.035 &0.70& 0.064 & 0.029 & 0.022&0.66\\
$ \Omega_{DE}$ &0.26 & 0.31 & 0.14&0.52& 0.24 & 0.30 & 0.10&0.30\\
$ \Omega_b$ & 0.047 & 0.047 & 0.030&0.80& 0.016 & 0.012 &0.0077&0.63\\
$ h$ & 0.11 & 0.12 & 0.072 &0.78& 0.089 & 0.11 & 0.066&0.90\\
$ w_0$ & 0.34 & 0.28 &0.19&0.77 &0.17 & 0.21 & 0.12&0.85 \\
$ w_a$ & 0.46 & 1.56 & 0.37&0.69 & 0.34 & 1.45 & 0.21&0.41\\
$ \sigma_8$& N/A&0.044&0.032&0.52& N/A & 0.023 & 0.017&0.52\\
$n_s $ & N/A & 0.61& 0.40&0.43& N/A & 0.27& 0.22&0.71\\
 \hline
\end{tabular}
\end{center}
\end{table*}

\subsection{Beyond the Total SZ Decrement}
\label{subsec:discussion_SZprofile}

Although large upcoming SZ surveys will not produce resolved
images of a large fraction of the detected clusters, it should still
be possible to go beyond measuring the overall SZ decrement, and to
obtain at least a rough constraint on its profile.  For example, for
SPT and ACT, the expected angular resolution is 1', while clusters
typically extend a few arc minutes. Given that our cluster model
specifies the radial structure of the cluster, it is interesting to
ask whether this additional information might help further constrain
parameters.  To see how large these improvements could be, we divided
every cluster into two annular regions: an inner part with radius 2',
and an outer part from $r=2'$ to the virial radius. We assumed the SZ
survey could independently measure $Y$ in both regions. The
measurement errors are assumed to be proportional to the square root
of their solid angles, and the sum in quadrature of the two errors is
fixed to be 1 mJy (to be consistent with our preceding calculations,
which adopted 1 mJy for the cluster as a whole). The Fisher matrix
forecast that use both $Y$ observables are given in Table
\ref{tbl:constraints4}, with and without cluster parameters
included. The columns ${\cal R}$ list the ratio of the new constraints
to their corresponding old values from Tables~\ref{tbl:constraints1}
and \ref{tbl:constraints2}. As these ratios reveal, even the very
crude measurement of the $Y$ profile can improve the constraints by a
factor of $\sim 2$.

\begin{table}
  \caption{Constraints on cosmological parameters when the solid area
of the cluster is divided into two annular regions, an inner part with
angular radius 2' and an outer part outside radius 2'.  The
constraints assume the $Y$ parameters in both regions are measurable
independently. The columns ${\cal R}$ list the ratio of these
constraints to their corresponding values from
Tables~\ref{tbl:constraints1} and \ref{tbl:constraints2} which use a
single $Y$ parameter.}
    \label{tbl:constraints4}
\begin{center}
\begin{tabular}{ c c c c c }
\hline \hline
Parameter & \multicolumn{2}{c}{6 parameters} & \multicolumn{2}{c}{21 parameters}\\
 & $\sigma$ & ${\cal R}$ & $\sigma$& ${\cal R}$\\
\hline
$ \Omega_m$ & 0.010&0.19&0.050&0.58\\
$ \Omega_{DE}$ &0.042&0.21&0.18&0.65\\
$ \Omega_b$ & 0.0040&0.34&0.036&0.45\\
$ h$ & 0.048&0.63&0.075&0.65\\
$ w_0$ & 0.013&0.36&0.21&0.41\\
$ w_a$ & 0.12&0.57&0.30&0.47\\
 \hline
\end{tabular}
\end{center}
\end{table}

\subsection{Spectroscopic Coverage}
\label{subsec:discussion_spectra}

A significant caveat, mentioned above, is that full spectroscopic
X-ray coverage of all SZ clusters is likely to be unavailable, since
the X--ray flux scales with redshift as $D_L^{-2}$, rather than
$D_A^{-2}$ as SZ flux.  For simplicity, we here study the effect of
partial X--ray data, by discarding all clusters beyond $z=1$ from the
scaling relations.  The results are presented in
Table~\ref{tbl:constraints5}, where again the columns ${\cal R}$ list
the ratio of these new constraints to their old values in Tables
\ref{tbl:constraints1} and \ref{tbl:constraints2}. It turns out that
the degradation is small, when cluster structure uncertainty is
neglected -- the largest difference is in $\Delta w_0$, whose error
increases from 0.037 to 0.05. When the cluster structure parameters
are included, the degradation is more severe (a factor of 1.7 for
$w_0$).  Given that the number of clusters beyond $z=1$ is only about
5\% of the total, and that they can bring down the constraints by a
factor of $\sim 2$, it could be worthwhile to conduct deep X-ray
followup measurements of these $\sim 300$ clusters at $z>1$.  We note
here for reference that the eROSITA deep survey is planned to cover
200 deg$^2$; acquiring $\sim 30\%$ temperatures out to $z=1$, and/or
to follow up on $\sim 300$ clusters at $z>1$ will be challenging, but
should be feasible in a large dedicated X--ray survey (e.g., Haiman et
al. 2005).  An alternative method may be to rely only on X--ray
imaging data: Mroczkowski et al. (2008) find that a joint analysis of
SZ and X-ray imaging data, for 3 clusters detected with the SZA
instrument, yields temperature profiles that are in good agreement
with spectroscopic X-ray measurements.  This may significantly ease
the requirement on the depth of the X--ray survey.

\begin{table}
  \caption{Constraints from scaling relations up to $z=1$ only.}
    \label{tbl:constraints5}
\begin{center}
\begin{tabular}{ c c c c c }
\hline \hline
Parameter & \multicolumn{2}{c}{6 parameters} & \multicolumn{2}{c}{21 parameters}\\
 & $\sigma$ & ${\cal R}$ & $\sigma$& ${\cal R}$\\
\hline
$ \Omega_m$ & 0.066&1.2&0.20&2.3\\
$ \Omega_{DE}$ &0.23&1.1&0.32&1.1\\
$ \Omega_b$ & 0.014&1.2&0.21&2.7\\
$ h$ & 0.081&1.1&0.13&1.1\\
$ w_0$ & 0.050&1.4&0.92&1.7\\
$ w_a$ & 0.23&1.1&1.11&1.8\\
 \hline
\end{tabular}
\end{center}
\end{table}

\subsection{Impact of Flat Universe Prior}
\label{subsec:discussion_flatprior}

We also note that most other forecasts in the literature tend to adopt
a flat universe prior.  When we impose this condition
($\Omega_m+\Omega_{DE}=1$), we find that our conclusions do not change
significantly.  In the idealized case without cluster parameters, we
find that the NC-alone constraint on $\Delta w_a$ is reduced from 1.4
to 0.47, and the SR-alone constraint on $\Delta w_0$ is reduced from
0.037 to 0.012.  These improvement, however are much less significant
when cluster structure uncertainty is taken into account (the NC-alone
constraint on $\Delta w_0$ is reduced from 1.61 to 1.0, and the
SR-alone constraint on $\Delta w_a$ is reduced from 0.53 to 0.51).

\subsection{Utilizing CMB Constraints}
\label{subsec:discussion_cmb}

We also computed the joint constraints on cosmological parameters from
the scaling relations, number counts and from the CMB temperature and
polarization anisotropy measurements by the upcoming {\it Planck}
satellite. Table~\ref{tbl:constraints7} presents these result. The
{\it Planck} Fisher matrix is adopted from Heavens et al. (2007).
This matrix assumes a flat universe, so we used the SR and NC matrices
with the same assumption, and we placed a prior of 0.1 on all cluster
parameters. Unsurprisingly, the addition of the SR+NC data improve
little over the {\it Planck}--alone constraints for most cosmological
parameters, except $w_0$, $w_a$ and $\sigma_8$. As is well known, the
{\it Planck} data alone for these 3 parameters suffer from severe
degeneracies (their degeneracy parameters $\cal{D}$, as defined in
Table \ref{tbl:constraints1}, are 161, 209 and 93 respectively).
Adding the cluster data causes significant improvements in these
parameters.  In fact, the improvements in the SR+NC+{\it Planck}
combination are significant compared either to SR+NC or to {\it
  Planck} alone; this shows that the improvements arise by breaking
degeneracy between the {\it Planck} and cluster dataset.  The joint
constraints on $w_0$ and $w_a$ improve to the interestingly tight
levels of 0.04 and 0.12, respectively.

\begin{table}
  \caption{Constraints including CMB anisotropy measurements by {\it Planck}. 
The constraints assume a flat
universe, and a prior of 0.1 is applied to all cluster
parameters. Only constraints on cosmological parameters are listed.}
    \label{tbl:constraints7}
\begin{center}
\begin{tabular}{ c c c c c}
\hline \hline
Parameter & SR & NC & Planck & Combined \\
\hline
$ \Omega_m$ & 0.059 & 0.029 &0.0023&0.0020\\
$ \Omega_b$ & 0.015 &0.012 &0.00069&0.00059\\
$ h$ & 0.080 &0.11 &0.0049&0.0042\\
$ w_0$ & 0.14 & 0.20 &0.35&0.04\\
$ w_a$ & 0.33 &0.52 &1.26&0.12\\
$\sigma_8$ &N/A& 0.023&0.074&0.0050\\
$n_s$& N/A & 0.26&0.0033&0.0022\\
 \hline
\end{tabular}
\end{center}
\end{table}

\subsection{Covariance Between Scaling Relations and Number Counts}
\label{subsec:discussion_covariance}

In our analysis, we have neglected correlations between the
constraints derived from different observables, i.e., the Fisher
matrices were simply added to obtain the joint constraints. Apart from
possible systematic measurement errors, the $Y$ vs. $T$ measurements
of two different clusters, in different regions of the sky, should
indeed be uncorrelated (except, perhaps in the rare cases of very
close neighbors physically affecting each other).  On the other hand,
it is not obvious whether correlations between the SR and NC
approaches are negligible.  For a given cluster, the $Y-M$ and the
$Y-T$ relations could well be correlated, through the underlying
physical origin of the scatter in these two relations.  Those clusters
with unusually high $Y$ values would then also be more likely to be
included in the detected sample. Indeed, in the limit that $M$ and $T$
are uniquely related without any scatter, deviations in $Y$ from the
expected $\langle Y\rangle$ will change the $Y$-$T$ relation and can
simultaneously affect the number counts (by moving a cluster to a
different $Y$-bin).

To see if this indeed introduces a significant correlation between the
SR and NC constraints, we performed a suite of Monte Carlo
calculations, in which 500,000 random realizations of a mock cluster
catalog were generated. At the assumed redshift $z=0.5$, each mock
catalog contains $\sim 2500$ clusters, drawn randomly from the
underlying mass function above the mass limit of $10^{14}~{\msun}$.
This limit corresponds to an SZ signal of $1\sigma$ at $z=0.5$,
sufficiently below the $5\sigma$ detection threshold that clusters
with masses below this limit have negligible chance to scatter above
the detection threshold.  The total number of clusters in each
realization of the mock catalog is drawn from a Poisson distribution
with a mean of 2,500. The number of detected clusters is then $\sim
400$, which is roughly the number of clusters in the redshift bin
$0.45<z<0.5$ in our fiducial model, assuming the survey parameters
given in \S~\ref{subsec:parameters}.

Using our cluster model, we assign a temperature and a $Y$ parameter
to each cluster, based on its mass. A 10\% intrinsic scatter and a
$1\sigma$ measurement uncertainty, which are drawn independently from
Gaussian distributions, are also added to the $Y$ parameter, but no
scatter is added to the temperature. By assigning scatter only to the
$Y$ parameter (and not to $T$), the $Y-M$ and $Y-T$ relations are
fully correlated, so that any resulting correlation between the SR and
NC constraints will be overestimated.  Maximum correlation between the
two approaches is achieved by anti-correlating $Y-M$ and $Y-T$
relations, however, such anti-correlation does not seem physically
realistic.  We have assigned a Fisher matrix to each single cluster in
the SR approach, and these Fisher matrices might have different
correlation strength to the NC approach. For example, scaling relation
Fisher matrix of a cluster with $Y$ just above the detection threshold
seems more correlated to NC approach than average, because scatter of
$Y$ can easily make it undetected and therefore change the number of
detected clusters. So to be exact, we need consider correlation for
each single SR Fisher matrix, which is a daunting task. Instead, In
order to reduce the dimensions of the covariance matrix, we binned the
SR observables into the same $Y$ bins used for the number counts (this
does not significantly change the SC constraints since the $Y-T$
relation does not have small-scale features that would be missed by
this binning).  We then studied the correlation between cluster number
$N$ and the binned SR observable $\bar{S}\equiv \langle
Y/Y_{0}(T)\rangle_{\rm bin}$, where $Y_{0}(T)$ is the $Y$ parameter
computed from the cluster model (without scatter), and the subscript
``bin'' indicates that the averaging is over all clusters in a $Y$ bin
(as opposed to over Monte Carlo realizations).

As a result of the intrinsic scatter and the measurement uncertainty,
$\bar{S}$ generally deviates from unity.  Using the 500,000 Monte
Carlo realizations, we computed the following three types of
correlation coefficients:
\begin{eqnarray}
r_{N_iN_j}=\frac{{\rm Cov}(N_i,N_j)}{\sigma_{N_i}\sigma_{N_j}},\nonumber\\
r_{N_i\bar{S}_j}=\frac{{\rm Cov}(N_i,\bar{S}_j)}{\sigma_{Ni}\sigma_{\bar{S}_j}},\nonumber\\
r_{\bar{S}_i\bar{S}_j}=\frac{{\rm Cov}(\bar{S}_i,\bar{S}_j)}{\sigma_{\bar{S}_i}\sigma_{\bar{S}_j}},\nonumber
\end{eqnarray}
where $i$ and $j$ refer to the $Y$-bin indices and, e.g., ${\rm
Cov}(N_i,N_j)\equiv\langle
(N_i-\bar{N_i})(N_j-\bar{N_j})\rangle^{1/2}$, where bar denotes
averaging within the $Y$-bin, and the other correlation coefficients
are defined similarly. First, without any scatter on $Y$, we checked
that no correlation is introduced artificially in the treatment
outlined above (i.e. all coefficients $r$ are zero). When we include
the scatter on $Y$, we find all three types of correlations
coefficient are a few $\times 10^{-3}$, which is still consistent with
zero within the uncertainty of the calculation.

While this result may be surprising, the lack of correlations is
explained by the fact that the $Y$--values are still drawn
independently for each cluster, even after the scatter is included. In
particular, moving clusters across adjacent $Y$-bins clearly
introduces correlations in the number of excess clusters {\it relative
to the no--scatter case}. The inclusion of scatter, however, also
changes the mean number of clusters in each $Y$--bin, and the $Y/T$
ratio in these bins.  It is the excursions around these modified mean
values that we find to be essentially uncorrelated.  In the Appendix,
we use a simplified toy model, in which we show that the covariance
between neighboring bins ${\rm Cov}(N_i,N_j)$ is strictly zero; one
can similarly show ${\rm Cov}(N_i,\bar{S}_j)=0$ and ${\rm
Cov}(\bar{S}_i,\bar{S}_j)=0$.

\subsection{The Importance of the Virial Overdensity}
\label{subsec:discussion_rhovir}

Finally, as mentioned above, there is an apparent tension between our
results, and the conclusions reached recently by Aghanim et
al. (2008), who find that the scaling relations are little affected by
changes in the dark--energy equation of state.  However, we believe
these two results are not, in fact, in contradiction.  We find that
the dark energy parameters $w_0$ and $w_a$ indeed have only a small
direct effect on the scaling relation.  For instance, we find that the
value of $Y$ for a cluster with a fixed temperature, at $z=0.2$,
increases by about 4\% when changing $w_0=-1$ to $w_0=-1.2$ (see
Table~\ref{tbl:dependence1}).  This is consistent with Figure 4 in
Aghanim et al., which shows that the normalization of the $Y-T$
relation changes by a few percent over the $-1.2 < w_0 < -0.8$ range,
with the largest increase for the smallest value $w_0=-1.2$ consistent
with $4\%$.  We note that Aghanim et al. also include a
cosmology--dependent factor, which is a power--law in the normalized
Hubble parameter $E(z)$, in their definition of $Y$. At low redshift,
this factor is $\sim 1$, and does not drive the $w_0$--sensitivity.
At higher redshifts, this factor drives the $w_0$--sensitivity; if we
were to scale out this factor from the definition of $Y$, at $z=1.5$
we would predict a $\sim 3\%$ increase in $Y$ when changing $w_0=-1$
to $w_0=-1.2$ which is again consistent with the change seen in Figure
4 in Aghanim et al. (their middle panel).  The constraints we obtain
here on $w_0$ and $w_a$ derive in large part from this modest
dependence of the scaling relations on the dark energy equation of
state, and owe much to the large number of clusters and relatively
weak degeneracies. Our results are also explicitly based on the
simulations by Kuhlen et al. (2005) on how dark energy affect the
virial overdensity. Although there is no clear sign of any
disagreement, it would be worthwhile to explicitly check the
consistency between the $w_0$--dependence of the virial overdensity
between the two simulations.

%%%%%%%%%%%%%%%%%%%%%%%%%%%%%%%%%%%%%%%%%%%%%%%%%%%%%%%%%%%%%%%%%%%%%%%%%%%%%%%%
\section{Conclusions}
\label{sec:conclusion}
%%%%%%%%%%%%%%%%%%%%%%%%%%%%%%%%%%%%%%%%%%%%%%%%%%%%%%%%%%%%%%%%%%%%%%%%%%%%%%%%

In this paper, we studied the utility of the scaling relation between
the Sunyaev--Zeldovich decrement $Y$ and temperature $T$ of galaxy
clusters; in particular, the constraint that this relation may place
on cosmological and on cluster structural parameters. A
phenomenological cluster model is adopted, which has 15 free
parameters to describe cluster structure, and its dependence on mass
and redshift. We demonstrated that this model fits available cluster
observations, including the temperature profile outside the core. We
then used this model to forecast constraints that could become
available from a future survey, containing several thousand clusters.

Our basic result is that the scaling relations have a statistical
constraining power on cosmological parameters comparable to those from
cluster number counts, even after we marginalize over the cluster
parameter uncertainties. We investigated where the cosmology
sensitivity in the scaling relation comes from, and found that the
constraints are driven by different physics for different parameters.
The constraints on $\Omega_m$ and $\Omega_b$ arise mainly through the
gas fraction $f_g$; the constraints on $\Omega_{DE}$ and $h$ are
predominantly through $D_A$, and the constraints on $w_0$ and $w_a$
are driven by the characteristic virial overdensity
$\Delta_{\mathrm{vir}}$ at low redshifts, but by $D_A$ at high
redshifts.

The scaling relation constraints have significant degeneracies.  The
most significant of these is between $\Omega_b$ and $\Omega_m$,
whereas the parameters suffering the least degeneracy are $w_0$ and
$w_a$. These dark energy equation--of--state parameters have
statistical errors from the scaling relations that are somewhat
tighter than from the number counts.  Combining the scaling relation
with the number counts, including multiple $Y$--bins, and combining
cluster data with expected CMB temperature and polarization
measurements by {\it Planck} all help in breaking parameter
degeneracies.  In a model that uses 6,800 clusters, and combines the
SR+NC+{\it Planck} data, and assumes a prior of 10\% on the 15 cluster
model parameters, we find tight constraints on the dark energy
equation of state parameters, $\Delta w_0=0.04$ and $\Delta w_a=0.12$.

The most significant caveat to our conclusions is that we assumed a
particular parameterization of the cluster model. Strictly speaking,
our numerical results are valid only within the confines of this
specific model.  However, our results do indicate, more generally,
that there will be significant statistical constraining power in the
scaling relations data, and that there is also sufficient cosmological
sensitivity in the scaling relations to be relevant as a cosmology
probe.  The latter point is the main new conclusion of this
work. Indeed, the cosmology--dependencies that we find drives the SR
constraints should be relatively robust features: $D_A$, $\Delta_{\rm
vir}$, and $f_{\rm g}$ are likely to depend on cosmology, in any
reasonable cluster model, in the way we exploited here.  Indeed, the
likely outcome of the analysis of any actual cluster survey data is
that it will force us to adopt some other, physically motivated
parametric description of cluster structure, and the parameters of
that new model will then have to be constrained, together with the
cosmological parameters.

We have studied various other explicit caveats, such as the impact of
partial X--ray coverage of the SZ sample, or uncertain scatter and
completeness, or departures from simple power--laws in the dependence
of the cluster parameters on mass and redshift.  We generally find
that the constrain loosen, as expected, but remain appreciable.
Overall, our work suggests that explicit scaling relations do not, by
themselves, strongly constrain the highly degenerate cluster
parameters, but that they should be a useful component in extracting
cosmological information from large future cluster surveys.

\vspace{-0.5\baselineskip}

\section*{Acknowledgments}

We thank Wenjuan Fang, Sheng Wang, Neelima Sehgal, Gil Holder, Amber
Miller, and Greg Bryan for many useful discussions, and the
  anonymous referee for useful comments. ZH acknowledges support by
the Pol\'anyi Program of the Hungarian National Office of Technology
and by the NSF grant AST-05-07161.  LV acknowledges the support of
Marie Curie International Reintegration Grant
FP7-PEOPLE-2007-4-3-IRGn202182.

%\vspace{-2\baselineskip}

\appendix
\section{Scatter and the (Lack of) Correlations}
\label{app:derivation}

As mentioned in \S~\ref{subsec:discussion_covariance} in the text, in
the limit that $M$ and $T$ are uniquely related without any scatter,
deviations in $Y$ around its mean value $\langle Y\rangle$ can
simultaneously change the $Y$-$T$ relation and affect the number
counts (by moving a cluster to a different $Y$-bin).  It is not
a--priori obvious whether this introduces correlations between the
number counts in different $Y$ bins, and/or cross--correlations
between the scaling relations and the number counts.  However, in
\S~\ref{subsec:discussion_covariance}, we performed Monte Carlo
simulations and found that that the correlations and
cross--correlations are both negligibly small.  In this Appendix, we
illustrate how this (lack of) correlation can arise, using a
simplified toy model.

Consider an experiment in which there are two bins, initially with a
random number $n_1$ and $n_2$ of objects in each bin, drawn
independently from Poisson distributions with means of $\bar{n_1}$ and
$\bar{n_2}$. Subsequently, each object in bin $\#1$ is either
relocated to bin $\#2$ with a probability $p$, or left in bin $\#1$,
with a probability of $(1-p)$, after which there are $N_1$ and $N_2$
objects in the two bins. To simplify the mathematics below, we do not
consider relocating objects from bin $\#2$ to bin $\#1$; this does not
affect the essence of the argument. The covariance between $N_1$ and
$N_2$ can be written as
\begin{eqnarray}
\label{eqn:cov}
{\rm Cov}(N_1, N_2)&=&{\rm Cov}(N_1, N_2-n_2+n_2)\\\nonumber
&=&{\rm Cov}(N_1, \Delta n+n_2)\\\nonumber
&=&{\rm Cov}(N_1, \Delta n)+{\rm Cov}(N_1, n_2),\\\nonumber
&=&{\rm Cov}(N_1, \Delta n)+{\rm Cov}(n_1-\Delta n, n_2),\\\nonumber
&=&{\rm Cov}(N_1, \Delta n)-{\rm Cov}(\Delta n, n_2),
\end{eqnarray}
where $\Delta n\equiv N_2-n_2=n_1-N_1$ is the number of objects
relocated from bin $\#1$ to bin $\#2$, and in the last step, we have
used ${\rm Cov}(n_1, n_2)=0$, which is true by construction.

In order to compute the last two terms on the right hand side, we
first need to know the probability distribution of $\Delta n$,
\begin{eqnarray}
\label{eqn:pdeltan}
P(\Delta n)=\sum_{n_1=\Delta n}^{\infty}P(\Delta n| n_1)P(n_1).
\end{eqnarray}
By definition, $n_1$ is drawn from Poisson distribution,
\begin{eqnarray}
\label{eqn:pn1}
P(n_1)=Pois(n_1;\bar{n_1})=\frac{\bar{n_1}^{n_1}e^{-\bar{n_1}}}{n_1!},
\end{eqnarray}
and, given $n_1$, $\Delta n$ follows a binomial distribution,
\begin{eqnarray}
\label{eqn:bino}
P(\Delta n| n_1)&=&Bino(\Delta n;n_1,p)\\\nonumber
&=&\frac{n_1!}{\Delta n!(n_1-\Delta n!)}p^{\Delta n}(1-p)^{n_1-\Delta n}.
\end{eqnarray}
Substituting equations \ref{eqn:pn1} and \ref{eqn:bino} to equation
 \ref{eqn:pdeltan}, we find
\begin{eqnarray}
\label{eqn:pdn2}
P(\Delta n)&=&\\\nonumber
=&&\sum_{n_1=\Delta
  n}^{\infty}\frac{\bar{n_1}^{n_1}e^{-\bar{n_1}}}{n_1!}
\frac{n_1!}{\Delta n!(n_1-\Delta n!)}p^{\Delta n}(1-p)^{n_1-\Delta
  n}\\\nonumber
=&&\sum_{n_1=\Delta n}^{\infty}\frac{\bar{n_1}^{n_1}e^{-\bar{n_1}}}{\Delta n!(n_1-\Delta n!)}p^{\Delta n}(1-p)^{n_1-\Delta
  n}.
\end{eqnarray}
Substituting $n_1=N_1+\Delta n$, and moving outside the summation all
factors not containing $N_1$, equation \ref{eqn:pdn2} is simplified to
\begin{eqnarray}
\label{eqn:pdn3}
P(\Delta n)=\frac{e^{\bar{n_1}}(\bar{n_1}p)^{\Delta n}}{\Delta n !}\sum_{N=0}^{\infty} \frac{(\bar{n_1}(1-p))^{N_1}}{N_1!}.
\end{eqnarray}
Equation \ref{eqn:pdn3} can be simplified further by noting that the
summation on the right hand side is the Taylor expansion of
$e^{\bar{n_1}(1-p)}$. After this simplification, we obtain the final
expression of $P(\Delta n)$,
\begin{eqnarray}
P(\Delta n)=\frac{(p\bar{n_1})^{\Delta n}e^{-p\bar{n_1}}}{\Delta n
  !}=Pois(\Delta n; p\bar{n_1}).
\end{eqnarray}
This corresponds to the intuitive result that $\Delta n$ follows a
Poisson distribution with a mean of $p\bar{n_1}$.  It can be shown
similarly that $P(N_1)=Pois(N_1;(1-p)\bar{n_1})$ and
$P(N_2)=Pois(N_2;\bar{n_2}+p\bar{n_1})$. ${\rm Cov}(N_1, \Delta n)$
can now be computed by considering the variance of $n_1$,
\begin{eqnarray}
\label{eqn:pdn4}
{\rm Var}(n_1)&=&{\rm Var}(N_1+\Delta n)\\\nonumber
&=&{\rm Var}(N_1)+{\rm Var}(\Delta n)+2{\rm Cov}(N_1,\Delta n).
\end{eqnarray}
Since $n_1$, $N_1$ and $\Delta n$ all follow Poisson distributions,
their variances are equal to their expectation values, which are
$\bar{n_1}$,$(1-p)\bar{n_1}$ and $p\bar{n_1}$, respectively. With this
information, it follows from equations \ref{eqn:pdeltan} and
\ref{eqn:pdn4} that
\begin{eqnarray}
{\rm Cov}(N_1, \Delta n)=0.
\end{eqnarray}
Similarly, ${\rm Cov}(\Delta n, n_2)=0$ follows by considering
the variance
\begin{eqnarray}
\label{eqn:pdn4}
{\rm Var}(N_2)&=&{\rm Var}(n_2+\Delta n)\\\nonumber &=&{\rm
Var}(n_2)+{\rm Var}(\Delta n)+2{\rm Cov}(\Delta n, n_2),
\end{eqnarray}
and noting ${\rm Var}(N_2)=\bar{n_2}+p\bar{n_1}$, ${\rm
 Var}(n_2)=\bar{n_2}$, and ${\rm Var}(\Delta n)=p\bar{n_1}$.  This
 concludes the proof that ${\rm Cov}(N_1, N_2)= {\rm Cov}(N_1, \Delta
 n)+ {\rm Cov}(\Delta n, n_2)=0$.

The derivation above can be generalized to show ${\rm Cov}(N_i,N_j)=0$
when $i\ne j$ in the case of multiple $Y$ bins. Similar to equation
\ref{eqn:cov}, ${\rm Cov}(N_i,N_j)$ could be decomposed as
\begin{eqnarray}
\label{eqn:cov2}
{\rm Cov}(N_i,N_j)&=&{\rm Cov}(\sum_{l}\Delta n_{l\rightarrow
  i},\sum_{m}\Delta n_{m\rightarrow j})\\\nonumber &=&\sum_{l,m}{\rm
  Cov}(\Delta n_{l\rightarrow i}, \Delta n_{m\rightarrow j}),
\end{eqnarray}
where $n_{l\rightarrow i}$ is the number of clusters relocated from
bin $\#l$ to bin $\#i$ by the scatter. For $l\ne m$, ${\rm Cov}(\Delta
n_{l\rightarrow i}, \Delta n_{m\rightarrow j})=0$ because of the
independence between $n_l$ and $n_m$, the number counts in the absence
of scatter. To compute ${\rm Cov}(\Delta n_{l\rightarrow i}, \Delta
n_{l\rightarrow j})$, we first need the probability distributions of
$\Delta n_{l\rightarrow i}$, $\Delta n_{l\rightarrow j}$ and $\Delta
n_{l\rightarrow i,j}\equiv (\Delta n_{l\rightarrow i}+\Delta
n_{l\rightarrow j})$. Following a procedure similar to the case of two
bins above, we can show that these quantities all follow Poisson
distributions with expectation values of $p_{l\rightarrow
i}\bar{n_l}$, $p_{l\rightarrow j}\bar{n_l}$ and $p_{l\rightarrow
i,j}=(p_{l\rightarrow i}+p_{l\rightarrow j})\bar{n_l}$,
respectively. Here, $p_{l\rightarrow i}$ is the probability of a
cluster in bin $\#l$ being scattered into bin $\#i$. For the Gaussian
scatter assumed in this paper,
\begin{eqnarray}
p_{l\rightarrow i}=\frac{\int_{M_{\rm min}(Y_l)}^{M_{\rm
      max}(Y_l)}g(Y_i,M)\frac{dn}{dM} dM}{\int_{M_{\rm min}(Y_l)}^{M_{\rm
      max}(Y_l)}\frac{dn}{dM} dM},
\end{eqnarray}
where $M_{\rm min}(Y_l)$ and $M_{\rm max}(Y_l)$ are the minimum and
maximum masses in bin $\#l$ in the absence of scatter, and $g(Y_i,M)$
was given in equation~(\ref{eqn:g}). By considering ${\rm Var}(\Delta
n_{l\rightarrow i,j})$, one can then show ${\rm Cov}(\Delta
n_{l\rightarrow i}, \Delta n_{l\rightarrow j})=0$; substituting back
into equation~(\ref{eqn:cov2}), we find ${\rm Cov}(N_i,N_j)=0$.

\end{document}